\documentclass{aa}
\usepackage[varg]{txfonts}
\usepackage{natbib,twoopt}
\usepackage{rotating}
\usepackage{multirow}

\renewcommand{\linenumbers}{}   
\usepackage{lineno}             
\nolinenumbers                  

\usepackage[colorlinks=true,     linkcolor=blue, citecolor=blue, filecolor=blue, urlcolor=blue]{hyperref}

\begin{document}



\title{Quantifying the scale of star formation across the Perseus spiral arm using young clusters around Cas~OB5}

\author{Alexis L. Quintana\inst{\ref{inst1}} \and Ignacio Negueruela\inst{\ref{inst1},\ref{inst2}}\and Sara R. Berlanas\inst{\ref{inst3}, \ref{inst4}}} 
\institute{Departamento de Física Aplicada, Facultad de Ciencias, Universidad de Alicante, Carretera de San Vicente s/n, 03690 San Vicente del Raspeig, Spain\label{inst1} \and
Instituto Universitario de Investigaci\'on Inform\'atica, Universidad de Alicante, San Vicente del Raspeig, Spain\label{inst2} \and Instituto de Astrofísica de Canarias, E-38 200 La Laguna, Tenerife, Spain\label{inst3} \and Departamento de Astrofísica, Universidad de La Laguna, E-38 205 La Laguna, Tenerife, Spain\label{inst4}}

\date{Received date 15 October 2024 /Accepted date 24 March 2025}

\abstract
{Cas~OB5 is an OB association located at a distance of 2.5--3 kpc that intercepts the Perseus spiral arm. It carries a moderate amount of reddening ($A_V \sim$ 2\,--\,3~mag) and contains several well-known open clusters within its boundaries, such as King~12, NGC~7788, and NGC~7790. The availability of modern clustering algorithms, together with \textit{Gaia} DR3 kinematics and complementary spectroscopic data, makes it a suitable site for studies of Galactic structure.}
{We seek to quantify the spatial scale of star formation in the spiral arms, using Cas~OB5 as a pilot target before extending our study to more distant and extinguished regions of the Galaxy.}{We selected 129\,695 candidate OBA stars in a 6x8 deg$^2$ region around Cas~OB5.   We applied a spectral energy distribution (SED) fitting process to this sample to derive the physical parameters. Through this process, we found 56\,379 OBA stars, which we then clustered using HDBSCAN.}
{We identified 17 open clusters inside this area, four of which appear to form a coherent structure that we identify as Cas~OB5. Nevertheless, our findings suggest that these clusters belong to two different age groups despite sharing a similar position and kinematics. Spectroscopic observations confirm the youth of NGC~7788 (10\,--\,15 Myr) compared to NGC~7790 ($110\pm15\:$Myr).}
{We have determined a spatial scale for star formation of a few tens of pc to a few hundreds of pc, comparing the clustered to the diffuse population of Cas~OB5 across this part of the Perseus arm. A spectroscopic analysis was required to complement the clustering algorithm, so that we could separate younger OCs (tracers of the spiral arm) from older ones. These results highlight the need to combine these techniques to fully disentangle the Milky Way structure.}


\titlerunning{Young clusters around Cas~OB5}
\authorrunning{Quintana, A.~L., et al.} 

\maketitle



\section{Introduction}
\label{introduction}

Star formation is a fundamental process that shapes the universe in all its vastness and complexity. In the Milky Way, it preferentially occurs within the spiral arms, where molecular clouds are more numerous (e.g. \citealt{Elmegreen2011}). The complexity of the physics involved, both at small and large scales (e.g. \citealt{Girichidis2020}), hinders a full understanding of this process. Notably, in spite of recent progress, such as the work presented by \citet{Pessa2021}, the spatial scale that sees star formation occurring in the spiral arms still needs to be fully quantified. 

A first step towards resolving this issue is to reliably trace the position of the Galactic spiral arms, for which open clusters and OB associations are suitable candidates \citep{Wright2020,CastroGinard2021,JoshiMalotra2023}. Due to their youth, these stellar groups will remain close to their environment. In addition, it is easier to determine a reliable distance for them than for individual stars (e.g. \citealt{QuintanaWright2022}).

The availability of large-scale astronomical surveys, alongside modern computing techniques, allows for the derivation of more accurate cluster membership than ever. In particular, the \textit{Gaia} mission \citep{Gaia}, with its last data release mapping the 3D position of $\sim$1.5 billion astronomical sources with astrometry \citep{GaiaDR3}, has revolutionised the field of star clusters \citep{CantatGaudin2022,CantatGaudinCasamiquela2024}, enabling the compilation of the most homogeneous and comprehensive censuses of open clusters (OCs) ever realised  \citep[e.g.][]{CantatGaudinAnders2020,HuntReffert2023}. While less dense, OB associations have also been extensively revisited \citep{Wright2023}, notably thanks to their compact subgroups and/or OCs contained within their boundaries \citep[e.g.][]{Kounkel2018,Squicciarini2021,Ratzenbock2023}.

For this work, we focus on the Cassiopeia constellation. It is a region of massive star formation, filled with OCs \citep[e.g.][]{Frincha2008,Wu2009}, together with emission nebulae and star-forming regions \citep[e.g.][]{Kun2008}, hosting the prominent supernova remnants Cassiopeia A and 3C~10 \citep[e.g.][]{Ruiz2019,Vink2022}. Ten OB associations have been identified in this area \citep{Reddish1961,Humphreys1978}, believed to trace larger structures within the region alongside OCs. For example, both Cas~OB1 and Cas~OB7 have been proposed to belong to the putative Cassiopeia-Perseus family, included within the spiral arms at $\sim$2 kpc and with a diameter of $\sim$600~pc \citep{DelaFuenteMarcos2009}. Furthermore, Cas~OB7 may undergo sequential star formation together with Cas~OB4 and Cas~OB5: given the very similar distances proposed for all of them \citep{Humphreys1978}, it has even been suggested that they could form a single, larger structure \citep{Velasco}.

The Cassiopeia region thus offers a very favourable setting to study the scale of star formation, and in this paper, we have chosen to specifically analyse the area of the catalogued association Cas~OB5. Previously thought to be located $\sim$2.1~kpc away from the Sun \citep{MelnikDambis2009}, more recent research has placed it further away, between 2.5 and 3~kpc \citep{SaltovetsMcSwain2024}. \citet{Chentsov2020} even suggested the presence of a background population beyond 4~kpc. Cas~OB5 exhibits a moderate level of extinction ($A_V =$ 2\,--\,3~mag) and (more critically) it intercepts the Perseus arm \citep{DelaFuenteMarcos2009,MarcoNegueruela2016}. This makes Cas~OB5 an ideal target to apply our techniques at higher distances than in previous works. This study is meant to serve as a first step before potentially exploring more obscured regions, such as those between $l = 70\degr$ and $90\degr$, or more distant ones, such as those beyond $l >140\degr$, where there are no known tracers of the Perseus arm \citep{MarcoNegueruela2017}.

Cas~OB5 is believed to power the H\,{\sc ii} region Sh2-173 with an age of 0.6\,--\,1~Myr \citep{Cichowolski2009}. However, the presence of supergiants and hypergiants in Cas~OB5 \citep{Humphreys1978,Bartaya1994,Kusuno2013} suggests that this association may rather be 10\,--\,30~\ Myr old. Cas~OB5 harbours OCs of various ages: the three most prominent ones, King~12, NGC~7788, and NGC~7790, have age estimates of $20 \pm8$, $20^{+9}_{-8}$ and $67^{+36}_{-25}$ Myr, respectively, in the catalogue of \citet{HuntReffert2024}. While OB associations are expected to exhibit some age spreads \citep{Wright2023}, notably between their subgroups \citep{PecautMamajek2016}, this wide range of ages motivates a deeper analysis to fully disentangle the structure of the region. Moreover, OB associations can have various configuration, ranging from massive central clusters surrounded by a halo, as in Per~OB1 \citep{deBurgos20}, to scattered populations of OB stars with low concentration levels and a high degree of substructure, as in Cyg~OB2 \citep{Wright16}. In this regard, Cas~OB5 exhibits an intermediate configuration, with several moderately sized clusters embedded in a dispersed population, which renders it a compelling subject for study. 

This paper is structured as follows. In Section \ref{identification}, we describe our selection of candidate OBA stars around Cas~OB5, to which we applied the spectral energy distribution (SED) fitting to characterise them. In Section \ref{clusteringanalysis}, we explain how we grouped these stars into open clusters, which we then compared with catalogued clusters and analysed their individual members. In Section \ref{discussion}, we discuss our results and we contrast the kinematic signature of these clusters with their age, using the youngest ones to quantify the spatial scale of star formation in this section of the Perseus arm. Finally, we present our conclusions in Section \ref{conclusion}.

\section{Identification of OBA stars}
\label{identification}

In this section, we outline how we selected a sample of candidate OBA stars within our region of interest (Section \ref{data}), which we physically characterised using an SED fitting tool (Section \ref{SEDfitter}). This allowed us to produce a catalogue of OBA stars with estimated physical parameters (Section \ref{genresults}).

\subsection{Data and selection process}
 \label{data}
 
For this work, we aim to find the most massive members of the open clusters related to Cas~OB5. To this end, we started by querying \textit{Gaia} DR3 sources \citep{GaiaDR3} within the region defined by $l$ between $114\degr$ and $120\degr$, $b$ between $-4\degr$ and $4\degr$, and $d$ between 1.5 and 4~kpc, where $d$ is the geometric distance from \citet{BailerJonesGaiaDR3}. This region was chosen based on the position of the members of the historical Cas~OB5 association \citep{Humphreys1978} and accounting for measurement uncertainties (notably in \textit{Gaia} DR3 parallaxes).

We corrected \textit{Gaia} DR3 photometry and astrometry following the methods of \citet{Maiz2018}, \citet{Maiz2021}, and \citet{Maiz2022}. This allowed us to apply a more liberal threshold on the renormalised unit weight error (RUWE), setting it at RUWE\,$<8$. Furthermore, we only kept sources with  $|\frac{\varpi}{\sigma_{\varpi}}|> 2$, where $\varpi$ and $\sigma_{\varpi}$, respectively, correspond to the corrected \textit{Gaia} DR3 parallax and its uncertainty. 

We subsequently used the ICRS coordinates at epoch J2000 to crossmatch, with a $1\arcsec$ separation theshold, this \textit{Gaia} DR3 catalogue with 2MASS\footnote{2 Micron All Sky Survey} \citep[from which we exploited the $J$, $H,$ and $K_s$ photometric bands]{2MASS} and IGAPS\footnote{the INT Galactic Plane Survey} \citep[from which we incorporated the $g$, $r_U$, and $i$ bands]{Drew,Mongui} photometric bands. We also set several conditions for each photometric band to be considered valid and thus used in our SED fitting process. For the \textit{Gaia} photometry, we required $G_{\rm RP}$ and $G_{\rm BP}$ to fulfil $|C^*| < 3 \, \sigma_{C^{*}}$, where $C^*$ corresponds to the corrected excess flux factor in the $G_{\rm RP}$ and $G_{\rm BP}$ bands, while $\sigma_{C^{*}}$ stands for the power-law on the $G$ band with a chosen 3$\sigma$ level \citep{Riello}. Besides, we considered non-valid the $G$-band if $\sigma_G > 2 \, \sigma_{G_{\rm BP}} \sim 2 \, \sigma_{G_{\rm RP}}$ (as these sources are assumed to be partially unresolved binaries) and vice versa for the $G_{\rm BP}$ and $G_{\rm RP}$ bands \citep{Maiz2023}. Photometric bands from 2MASS with a bad quality photometric flag (`E', `F', `U', and `X') were excluded \citep{2MASS}. For IGAPS, we also removed the saturated photometric bands, along with those with an associated class that did not indicate a star or probable star \citep{Mongui}. 

We then applied a first, broad absolute magnitude cut in the near-infrared (NIR), as interstellar extinction has a lesser effect at these wavelengths \citep{Fitz2004}, retaining sources with $M_{Ks} < 2$ mag  if there was valid $K_s$-band photometry and using the $H$ and $J$ bands from 2MASS otherwise. A second cut was made on the $G$-band magnitude, with $M_G < 2$ mag. For each cut, we used the geometric distance from \citet{BailerJonesGaiaDR3} to convert from apparent to absolute magnitude, correcting for reddening only for the \textit{Gaia} absolute magnitude (as extinction in $K_{\mathrm{S}}$ is expected to be only a few tenths of a magnitude for the typical values in the area). For this correction, we applied an average conversion factor of $A_G = 0.843 \, A_V$, with $A_V = 2.742 \, \times \, 1.22 \, \times \, E(g-r)$, with the factor 0.843 stemming from \citet{Zhang2023}. The value of $E(g-r)$ was taken from the \textit{Bayestar2019} extinction map \citep{Bayestar} within a given sightline, after scaling it up by 22 \% from a calibration with the \textit{Gaia} DR3 extinction map \citep{Delchambre}, similarly to \citet{Quintana2023}. 

The threshold on the $G$-band absolute magnitude was based on the expected value for an A5\,V star, after  \citet{Mamajek}, with $A_G$ scaled up by 25 \% to account for the choice of an average conversion factor (in reality, there is a temperature dependence for the conversion from $A_G$ to $A_V$, see \citealt{Fouesnau2023}). Finally, we required at least one valid blue photometric band ($G$, $G_{\rm BP}$ or $g$). Within this final sample, 77 sources were identified as duplicates within $1\arcsec$, that is,\ they are separate sources in \textit{Gaia} DR3, but not in 2MASS. We thus decided to fit these sources without their 2MASS photometry, as we modelled them all as single stars. The process led to a final catalogue of 129\,695 candidate OBA stars and we applied the SED fitter to this catalogue.

\subsection{SED fitting}
\label{SEDfitter}

The method we applied to estimate physical parameters of our sample of candidate OBA stars is an SED fitting process, originally described in \citet{Quintana}, with an improved version outlined in \citet{Quintana2023}. In summary, we have sought to estimate the initial mass of a star, $\log(M/M_{\odot})$, alongside its fractional age Fr(Age) (based on its position on the evolutionary tracks from \citealt{Ekstrom2012}), its distance $d$ and $\ln(\rm f)$, an additional uncertainty incorporated to ensure convergence of $\chi^2$ \citep{Emcee}. 

To that end, we need an observed SED, composed of \textit{Gaia} DR3 parallax and the photometry from several optical and NIR surveys (here, \textit{Gaia} DR3, 2MASS, and IGAPS DR1), to which we added systematic uncertainties. An update over the version used in \citet{Quintana2023} is the use of distinct band passes for the \textit{Gaia} DR3 $G$-band for bright ($G < 13$~mag) and faint ($G > 13$~mag) sources, following the recalibration of \textit{Gaia} photometry \citep{MaizApellaniz2024}. We also needed a model SED, built with the BT-NextGen \citep{Asplund2009}, Kurucz \citep{Coelho2014}, and TMAP Grid 2 \citep{Werner2003} stellar atmosphere models (with effective temperatures ranging from 3000 to 50,000~K), the interpolation of stellar evolutionary models from \citet{Ekstrom2012}, and the \textit{Bayestar2019} 3D extinction map \citep{Bayestar} to derive the extinction as a function of distance (this extinction is scaled up by 22\% as in Section \ref{data}).

 We fit the model SED to the observed by using the \texttt{emcee} package from Python \citep{Emcee}. It is an MCMC process utilising a Bayesian modelling and based on a maximum-likelihood test. The parameter space is explored with 1000 walkers, 200 burn-in steps, and 200 iterations, except when $\ln(f) > -4$ or $\log(T_{\rm eff})P95 - \log(T_{\rm eff})P5 > 0.5$. In such cases, we ran 1000 supplementary burn-in steps and 200 additional iterations, until a convergence was reached or until 3000 supplementary burn-in iterations, above which the estimates cannot be significantly improved. The median of the posterior distribution was adopted for the final value of the parameters, while the 16th and 84th percentiles were used as the lower and upper uncertainties, respectively. Finally, $\log(T_{\rm eff})$ and  $\log(L/L_{\odot})$ are indirect products of the process, interpolated from $\log(M/M_{\odot})$ and Fr(Age) by using the evolutionary tracks of \citet{Ekstrom2012}.
    
\subsection{General results}
\label{genresults}

\begin{figure*}
    \centering
    \includegraphics[scale=0.09]{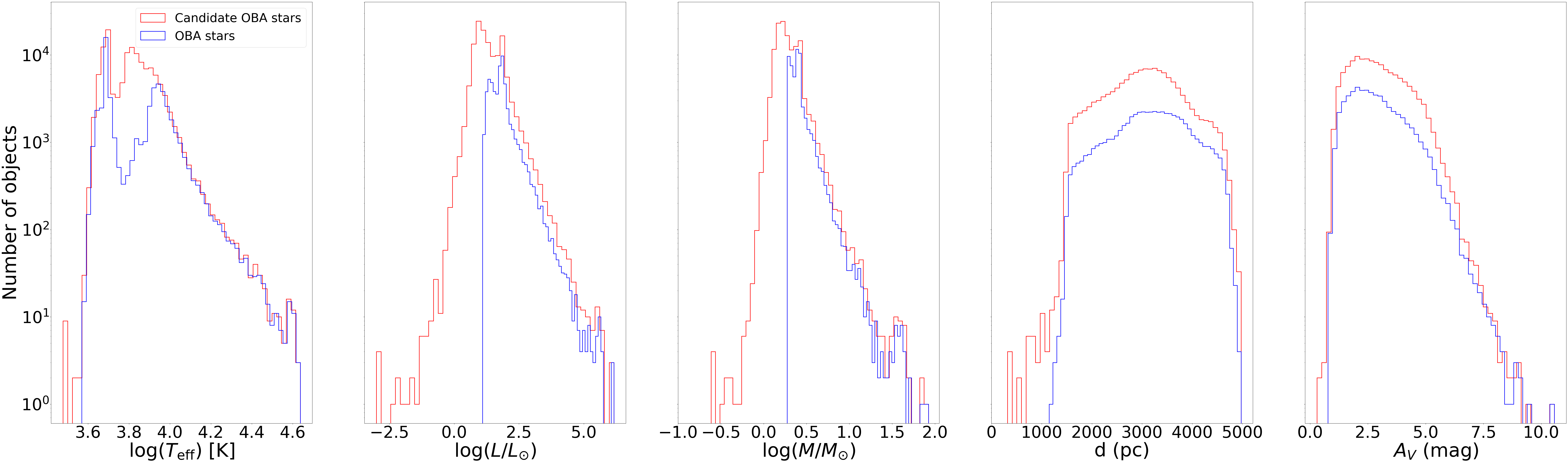}
    \caption{Histograms of the median SED-fitted parameters of the 129\,695 candidate OBA stars (in red) as well as of the 56\,379 SED-fitted OBA stars ($M > 1.88 \, M_{\odot})$, in blue) in the region.}
    \label{MedianParameters}
\end{figure*}

We applied the SED fitter described in Section \ref{SEDfitter} to the 129\,695 candidate OBA stars. The median SED-fitted parameters for these sources are displayed in Fig. \ref{MedianParameters}, revealing a bi-modal distribution of effective temperatures. The first peak corresponding to cool evolved stars (e.g. giants and red supergiants), while the second peak corresponds to main-sequence A-type stars. The median values of $\log(M/M_{\odot})$ and $\log(L/L_{\odot})$ of 0.25 and 1.22~dex, respectively, reinforce this picture.

Moreover, the median value of the SED-fitted distances is equal to 3.07 kpc, roughly consistent with the distance of Cas~OB5 following recent investigations \citep[e.g.][]{SaltovetsMcSwain2024}. Likewise, the median $A_V$ value is 2.77~mag, consistent with a moderate extinction encountered in this region. Notably, 7821 stars ($\sim14$ \%) of the sample are more distant than 4~kpc despite our initial selection criterion, based on the distances in \citet{BailerJonesGaiaDR3}. We attribute this discrepancy to the differences in the method of distance determination and the increasing uncertainties in \textit{Gaia} parallaxes at such distances. 

From the full sample, we selected  stars with $M > 1.88 M_{\odot}$, a threshold corresponding to the mass of a A5\,V star according to \citet{Mamajek}. This reduced the sample to a list of 56\,379 SED-fitted OBA stars, whose median parameters we have also displayed in Fig.~\ref{MedianParameters}.

\begin{figure}
    \centering
    \includegraphics[scale = 0.33]{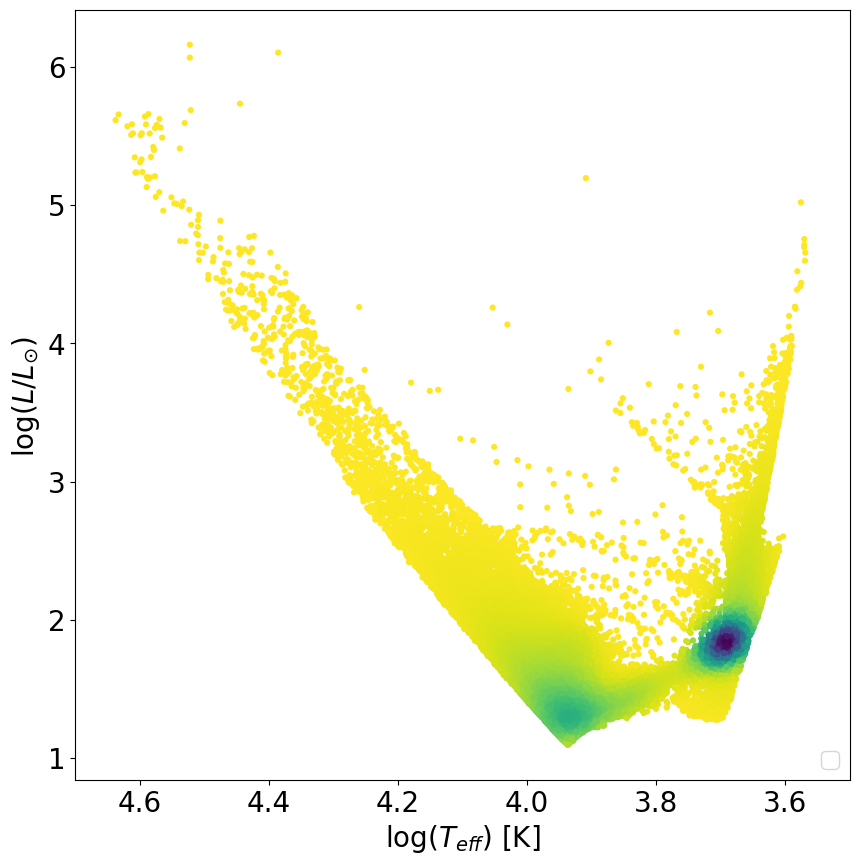}
    \caption{HR diagram for the 56,379 SED-fitted OBA stars in our sample, produced with their median effective temperature and luminosity, colour-coded by gaussian\_kde density.}
    \label{HRDMassiveStars}
\end{figure}

We  also made an HR diagram of these 56\,379 stars on the upper mass end of our sample, displayed in Fig.~\ref{HRDMassiveStars}. As expected, this subset is dominated by late B-type and early A-type stars, with the peak density on the lower part of the giant branch. While they might be sparser, some O-type stars and supergiants are also visible in this diagram. 

\section{Clustering analysis}
\label{clusteringanalysis}

In this section, we describe how we identify reliable clusters within the region (Section \ref{identificationclusters}), whose members are then compared with a homogeneous catalogue of open clusters (Section \ref{compother}). We proceeded to investigate our newly found clusters, specifically to assess their age and learn more about Cas~OB5 (Section \ref{investigation}), before studying their expansion (Section \ref{expansion}) and contrasting the clustered and diffuse population within the studied region (Section \ref{clusterediffuse}).
 
\subsection{Identification of open clusters}
\label{identificationclusters}

In recent years, the clustering algorithm HDBSCAN \citep{HDBSCAN} has been  used extensively to identify open clusters in \textit{Gaia} data  \citep[e.g.][]{HuntReffer2021,Tarricq2022,HuntReffert2023,Qin2023}, which motivates us to follow their approach. 
The main advantage of HDBSCAN over other clustering algorithms, such as DBSCAN \citep{DBSCAN}, is its ability to detect groupings at various densities. On the other side,  its overconfidence makes it necessary to reliably filter out false positives. We  chose a value of 10 for \texttt{min\_samples} and `leaf' for \texttt{cluster\_selection\_method}, as both are suited for the identification of stellar clusters. For the \texttt{min\_cluster\_size}, we have followed the method from \citet{HuntReffert2023} in trying several values. We found that a value below 20 increased the risk of finding unphysical clusters, whilst values above 50 become inappropriate as we are only using the most massive cluster members, which are less numerous. Therefore, we tried  \texttt{min\_cluster\_size} = 20, 30, 40, and 50.

We applied these four configurations of HSBCAN to the 56\,379 SED-fitted OBA stars, only keeping the clusters from a `lower configuration' (smaller value of \texttt{min\_cluster\_size}) if they were not found in a `bigger configuration'. The five parameters exploited for the clustering process were $X$, $Y$, $Z$, $V_l$, $V_b$, all converted from $l$, $b$, $d$, $\mu_l$, and $\mu_b$ (where $d$ corresponds to the median SED-fitted distance). We also normalised each parameter with respect to the parameter with the largest extent of this same unit. This means that $X$, $Y,$ and $Z$ were normalised with respect to $X$ (allowing us to remove any stretching along the line-of-sight), whereas $V_b$ was normalised with respect to $V_l$. This analysis resulted in the identification of 31 stellar groups within the region. 

\begin{figure}
    \centering
    \includegraphics[scale = 0.3]{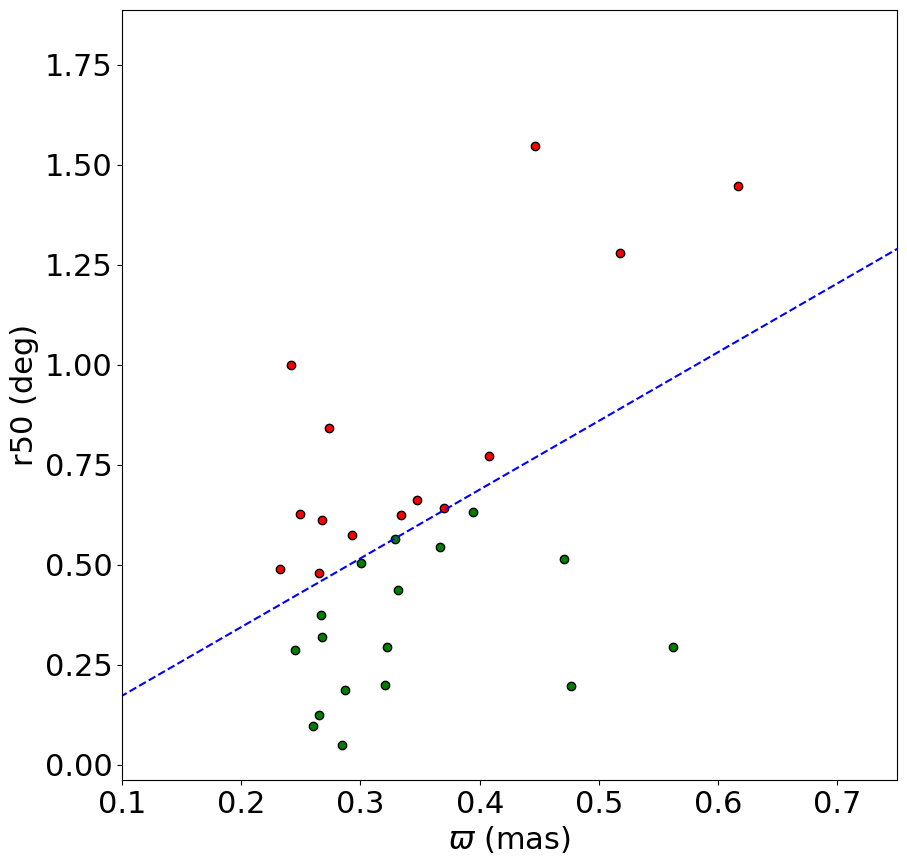}
    \caption{Radius containing half the cluster members as a function of their median parallax, where the blue dotted line corresponds to the equivalent radius of 30 pc. The kept groups are displayed as green dots, whilst the rejected ones are displayed as red dots.}
    \label{RadiusClusters}
\end{figure}

To reject false positives amongst these groupings, we  required that at least half of their members be contained within a radius equal to 30~pc. This threshold is higher than the value of 20~pc used by \citet{HuntReffert2023}; this is again because we are only considering the upper-mass end of cluster members, noting that their density would be lower than if we were considering all members, regardless of their mass. For this cut, we  followed the method from \citet{CantatGaudinAnders2020} and derived $r_{50}$ (in degrees), namely, the radius containing half the cluster members that we plotted as a function of the median parallax of the groups. Every cluster below or intercepting the line corresponding to a radius of 30~pc was kept, as illustrated in Fig.~\ref{RadiusClusters}. 

\begin{figure}
    \centering
    \includegraphics[scale =0.42]{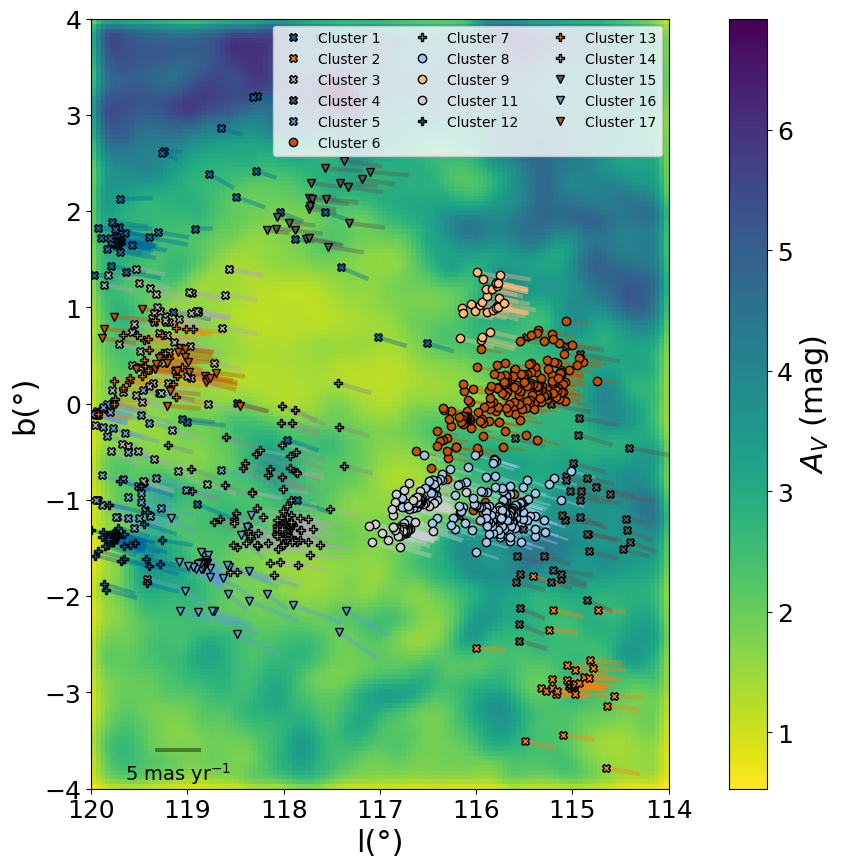}
    \caption{Galactic coordinates for the identified members of the 17 reliable open clusters around Cas~OB5, with their Galactic PMs shown as vectors. The background is the scaled extinction map from \citet{Bayestar}, smoothed over a step size of 0.05 deg. A scale for the length of the proper motion vectors is included on the bottom-left.}
    \label{GalCoordClusters}
\end{figure}

\begin{figure}
    \centering
    \includegraphics[scale =0.38]{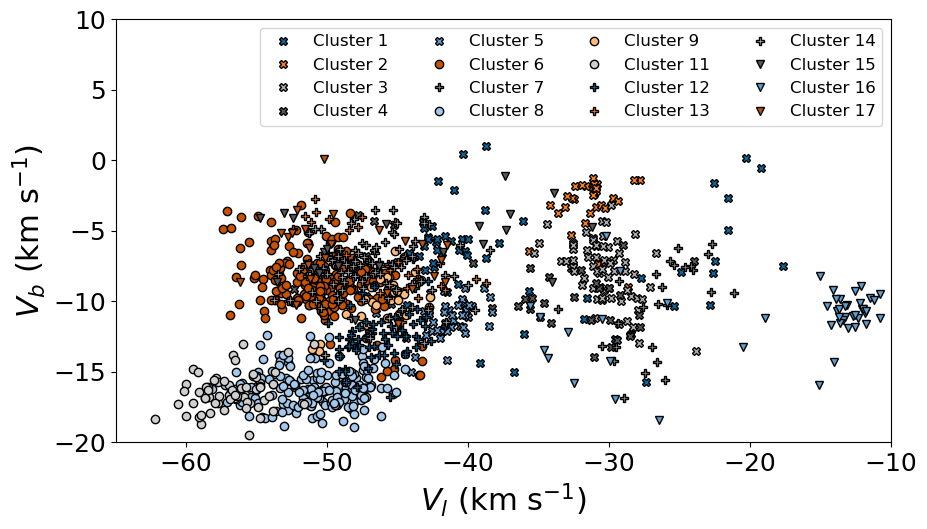}
    \caption{Galactic transverse velocities for the identified members of the 17 reliable open clusters around Cas~OB5.}
    \label{GalVelocityClusters}
\end{figure}

This criterion reduced the list of stellar groups to 17 reliable clusters, which we ordered by increasing median SED-fitted distance. We display their locations in Galactic coordinates in Fig.~\ref{GalCoordClusters} and their transverse velocities in Fig.~\ref{GalVelocityClusters}. We present their observed properties in Table \ref{TabClus}. In addition, we have shown their 3D distribution in Fig. \ref{3Dplotsclusters}.

Figure~\ref{GalCoordClusters} clearly reveals a sparsely populated area around $l \approx 117\degr$, where there are very few cluster members. This gap coincides with regions of very low extinction, and is thus real. Several clusters with lower Galactic longitude lie at distances consistent with that expected for Cas~OB5. At longitudes $\gtrsim118\degr$, there are clusters at many different distances, including two at distances that are comparable to that of Cas~OB5, but they are clearly disconnected to the association. 

\subsection{Comparison with other catalogue}
\label{compother}

The catalogue from \citet{HuntReffert2024} provides an ideal list for comparison, as it is the most comprehensive and homogeneous catalogue of open clusters produced with \textit{Gaia} DR3 data. It is an update of the catalogue from \citet{HuntReffert2023}, where the Jacobi radius was used to separate reliable, gravitationally bound open clusters from unbound moving groups. We  performed a cross-match within $1\arcsec$  between our cluster members and theirs. Out of the 1156 members of all our clusters, 489 have a counterpart in \citet{HuntReffert2024}. The results of this comparison are summarised in Table \ref{TabClus}.

Table \ref{TabClus} highlights the fact that while there are 1:1 correspondences between our clusters and those in  \citet{HuntReffert2024} (King 1, UBC 180, NGC 103, King 13, and Berkeley 60), some of their clusters are split between multiple of our clusters. Combined with the information from Figs.~\ref{GalCoordClusters} and~\ref{GalVelocityClusters}, we highlight four clusters that are related to Cas~OB5, whilst also considering their position and kinematics:

\begin{enumerate}
    \item Cluster 6, of median distance $2.81 \pm 0.08$~kpc, has a significant overlap with King~12 and Stock 17, alongside a few members in NGC~7788 and FSR 451.
    \item Cluster 8, of median distance $2.97 \pm 0.12$~kpc, has a significant overlap with FSR~451, but also includes a few members in Berkeley~58, NGC~7790 and NGC~7788.
    \item Cluster 9, of median distance $3.00 \pm 0.03$~kpc, has some members in Teutsch~23, King~21 and Negueruela~1.
    \item Cluster 11, of median distance $3.35 \pm 0.06$~kpc, has a significant overlap with NGC~7790 and Berkeley~58.
\end{enumerate}

It is worth noting that our clustering method gives more weight to the distance (derived from SED fitting) than the procedure used by \citet{HuntReffert2023} to separate the clusters. This contrast is particularly evident for NGC~7790 and Berkeley~58, which \citet{HuntReffert2024} placed at $3.01 \pm 0.02$ and $3.13 \pm 0.02$~kpc (see Table \ref{TabClus}), despite Cluster 11 being significantly more distant ($3.35 \pm 0.06$~kpc) and having many members in these two clusters. We discuss  this aspect further in Appendix~\ref{app:ngc}. In addition, our clusters with a 1:1 correspondence with \citet{HuntReffert2024} are either in the foreground or in the background with respect to Cas~OB5. Clusters related to Cas~OB5, on the other hand, are divided between many entities in the catalogue of \citet{HuntReffert2024}, warranting a deeper investigation (see Section \ref{investigation}).

\subsection{Investigation of selected clusters}
\label{investigation}

\begin{table*}
        \centering
        \caption{Properties of the main open clusters related to Cas~OB5 in \citet{HuntReffert2024} and our identified clusters, ordered by increasing median distance.}\label{OCsProperties}
        \renewcommand{\arraystretch}{1.5} 

        \begin{tabular}{ccccccc} 
                \hline \hline
                OC & OC (HR24) & N$_{\rm HM}$/N$_{\rm all}$ & $d$ (kpc) & Age (Myr) (HR24) & Age (Myr) (Others) & Age references \\
                \hline
         6, 8 &  NGC~7788 &  58/98 &$2.70^{+0.02}_{-0.01}$ & $20^{+9}_{-8}$ & 17\,--\,40 & [1], [2], [4]\,--\,[7], [8], [11], [12] \\
         6, 8 & FSR~451 &  118/186 & $2.80^{+0.02}_{-0.01}$ & $27^{+9}_{-15}$ &  11\,--\,14 & [5], [11], [12] \\
         6 & King~12 & 58/81 & $2.80^{+0.02}_{-0.01}$ & $20 \pm 8$ & 10\,--\,14 & [1]\,--\,[7], [11], [12] \\
         9 & King~21 &  82/153 & $2.89^{+0.01}_{-0.03}$ & $52^{+34}_{-17}$ & 10\,--\,117 & [1], [2], [5]\,--\,[7], [10]\,--\,[12] \\
        9 & Teutsch~23 & 17/19 & $2.89^{+0.04}_{-0.05}$ & $13^{+9}_{-5}$ & 8\,--\,64 & [2], [5]\,--\,[7], [11], \\
        6 & Stock~17 &  47/64 & $2.93^{+0.03}_{-0.02}$ & $12 \pm 5$ & 6\,--\,10 & [5]\,--\,[7], [11] \\
        9 &  Negueruela~1 &  11/11 & $2.95 \pm 0.04$ & $4 \pm 1$ & 3\,--\,17 & [5]\,--\,[7], [9], [11], [12]  \\
        8, 11 & Berkeley~58 & 91/173 & $3.01 \pm 0.02$ & $91^{+47}_{-36}$ & 158\,--\,372 &  [2], [5]\,--\,[8] [11], [12]\\
         8, 11 & NGC~7790 &  77/143 & $3.13 \pm 0.02$ & $67^{+36}_{-25}$ & 57\,--\,126  &  [2], [4]\,--\,[7], [11], [12]\\
        \hline
        \end{tabular}
\begin{minipage}{\linewidth} 
\vspace{0.25cm}
Note: N$_{\rm HM}$ stands for the number of cluster members in HR24 with a 50th percentile of the mass greater than 1.88 M$_{\odot}$, whilst N$_{\rm all}$ corresponds to the total number of cluster members. For comparison, the ages from other references have also been indicated on the second to last column, with the corresponding references written on the last column. References are [1]: \citet{Pandey1989}, [2]: \citet{Kharchenko2005}, [3]: \citet{Hancock2008}, [4]: \citet{Davidge2012}, [5]: \citet{Kharchenko2013}, [6]: \citet{Kharchenko2016}, [7]: \citet{LoktinPopova2017}, [8]: \citet{Bossini2019}, [9]: \citet{MonteiroDias2019}, [10]: \citet{Maurya2020}, [11]: \citet{Dias2021}, and [12]: \citet{Tarricq2021}.
\end{minipage}
\end{table*}

The properties of the main catalogued OCs related to our clusters and Cas~OB5 are summarised in Table~\ref{OCsProperties}. Our Cluster~6 is clearly the youngest, with every matching OC in \citet{HuntReffert2024} younger than 30 Myr, whilst Cluster~11 is the oldest, with all matching OC in \citet{HuntReffert2024} older than 50 Ma. Cluster~8 would thus act as a `bridge' between our youngest OC (Cluster 6) and our oldest OC (Cluster 11); whereas most of its members in \citet{HuntReffert2024} belong to the young OC FSR~451 (and a few in NGC~7788), it still has a sizeable number of members in NGC~7790 and Berkeley~58. The case of cluster 9 is more ambiguous, with King~21 noticeably older than Negueruela~1 and Teutsch~21 according to \citet{HuntReffert2024}, although age estimates vary significantly across different studies.

To further constrain the ages of our identified clusters, we have also looked into the available spectroscopic information on their individual members below. This investigation is complemented with spectroscopic observations of NGC~7788 and NGC~7790 in Appendix~\ref{app:ngc}, where we determined more reliable ages for these two clusters. 

\subsubsection{Cluster 6}

This cluster contains 6~Cas~A, which has been classified as an A3\,Ia star in \citet{Maiz2021}. In the same study, they also classified its companion as an O9.5\,II star, and estimated a distance of $d = 2.78^{+0.37}_{-0.29}$~kpc for this system, to be contrasted with our SED-fitted distance $2.68^{+1.02}_{-0.53}$~kpc with larger uncertainties (we used the \textit{Gaia} DR3 parallax while they relied on another method due to the lack of accurate \textit{Gaia} DR2 astrometry for this system). This cluster also contains BD~$+61\degr$2550, which has been classified as an O9.5\,II star in \citet{Suad2016}.

While some other cluster members have recorded spectral types in SIMBAD, going from mid-to-early B-type stars, these originate from much older references (e.g.\ \citealt{Brodskaya1953,Hardorp1959}) and do not have a recorded luminosity class, thereby making them less reliable. Nevertheless, given the  information above, the available spectroscopy for Cluster~6 confirms that it is a young cluster, although it also includes a few older stars. 

\subsubsection{Cluster 8}

This cluster contains no less than 13 members with a recorded spectral type and luminosity class in SIMBAD. Ten of them are early to mid B-type stars (either main-sequence or giants), whose spectral types have all been classified in \citet{Martin1972}, except for BD~$+60\degr$2631, whose spectral type (B0.5\,III) have been established through medium-resolution spectroscopy in \citet{NegueruelaMarco2003}. The three other stars are two classical Cepheid variables and one red supergiant; namely: V* CF Cas and V* CE Cas B of spectral type F8\,Ib \citep{Herbig1960} and V* TZ Cas of spectral type M3\,Iab \citep{KeenanMcNeil1989}.

With the above information, we can conclude that just like Cluster~6, Cluster~8 is young, although including a few old stars (as suggested by Table \ref{TabClus}: one-third of its members are in FSR~451 and $\sim10$ \% in NGC~7790 and Berkeley~58). Its most noticeable members are, however, classical Cepheid variable, which are inherently older. It is possible that our distance estimations from \textit{Gaia} DR3 parallaxes are underestimated for these cool supergiants, which may explain why they have been identified members of Cluster~8.

\subsubsection{Cluster 9}

This cluster only has a single member recorded with a spectral type in SIMBAD, LS~I~$+62\degr$24.  Given  it is just `B' without any reference, we cannot constrain the age of this cluster based on the available information on its individual members. 

\subsubsection{Cluster 11}

Only two members of this cluster have a recorded spectral type: TIC 326815525 of spectral type B8\,IV, as well as CG~Cas, a classical Cepheid variable (of spectral type F5) that is thus classed as an intermediate-mass  supergiant. Combined with the information from Table~\ref{TabClus}, this suggests that cluster 11 is an old cluster that includes a few young stars. 

\subsection{Expansion}
\label{expansion}

\begin{table*}
  \centering
   \caption{ Linear expansion gradients and corresponding expansion ages for our 17 identified OCs.  \label{ClusterExpansion}}
  \renewcommand{\arraystretch}{1.5}
    \begin{tabular}{|p{2cm}|p{2.5cm}|p{2.5cm}|p{2.5cm}|p{2.5cm}|}
    \hline
    \multirow{2}{2cm}{Cluster} & \multicolumn{2}{c|}{Linear expansion gradients (km s$^{-1}$ pc$^{-1}$)} & \multicolumn{2}{c|}{Linear expansion ages (Myr)}\\
    \cline{2-5}
    & $l$ &  $b$ & $l$ & $b$ \\
    \hline
    1 & $-0.205 \pm 0.001$ & $0.047 \pm 0.001$ & - &  $21 \pm 1$  \\ \hline
    2 & $0.014 \pm 0.006$ & $-0.006 \pm 0.001$ & $71^{+54}_{-21}$ &  - \\ \hline
    3 & $0.051^{+0.005}_{-0.003}$ & $0.077 \pm 0.001$ & $20^{+1}_{-2}$ & $13 \pm 1$ \\ \hline
    4 &  $0.081^{+0.006}_{-0.004}$ & $0.005 \pm 0.001$ & $12 \pm 1 $ &  $200^{+50}_{-33}$ \\ \hline
    5 & $0.022^{+0.007}_{-0.005}$ & $0.012 \pm 0.001$ & $45^{+14}_{-11}$ &  $83^{+7}_{-13}$ \\ \hline
    6 & $0.040 \pm 0.002$ & $0.057 \pm 0.001$ &  $25 \pm 1$ & $17 \pm 1$\\ \hline
    7 & $0.033 \pm 0.005$ & $0.011 \pm 0.002$ & $30^{+6}_{-4}$ & $91^{+20}_{-14}$ \\ \hline
    8 & $-0.008^{+0.002}_{-0.003}$ &  $0.011 \pm 0.001$ & - & $91^{+9}_{-7}$  \\ \hline
    9 & $-0.024^{+0.019}_{-0.031}$ & $0.062 \pm 0.005$ & - & $16^{+2}_{-1}$\\ \hline
    10 &  $0.024^{+0.005}_{-0.004}$ & $0.012 \pm 0.001$ & $42^{+8}_{-5}$ &  $83^{+8}_{-6}$ \\ \hline
    11 & $-0.011^{+0.011}_{-0.014}$ & $0.015 \pm 0.002$ & - & $67^{+10}_{-8}$  \\ \hline
    12 &  $0.014^{+0.014}_{-0.013}$ & $-0.006^{+0.003}_{-0.004}$ & - & - \\ \hline
    13 & $-0.041^{+0.012}_{-0.009}$ & $-0.001 \pm 0.002$ & - & - \\ \hline
    14 & $0.022 \pm 0.009$ & $-0.001 \pm 0.002$ &  $45^{+32}_{-13}$ &  -\\ \hline
    15 & $0.166^{+0.016}_{-0.013}$ & $0.012 \pm 0.003$ & $6 \pm 1$ & $83^{+18}_{-16}$ \\ \hline
    16 & $0.070^{+0.006}_{-0.004}$ & $0.021 \pm 0.002$ & $14 \pm 1$ & $48^{+5}_{-4}$    \\ \hline
    17 & $-0.070^{+0.014}_{-0.013}$ & $0.011 \pm 0.004$ & - &  $91^{+52}_{-24}$\\ \hline
  \end{tabular}
\end{table*}

\begin{figure}
    \centering
    \includegraphics[scale = 0.35]{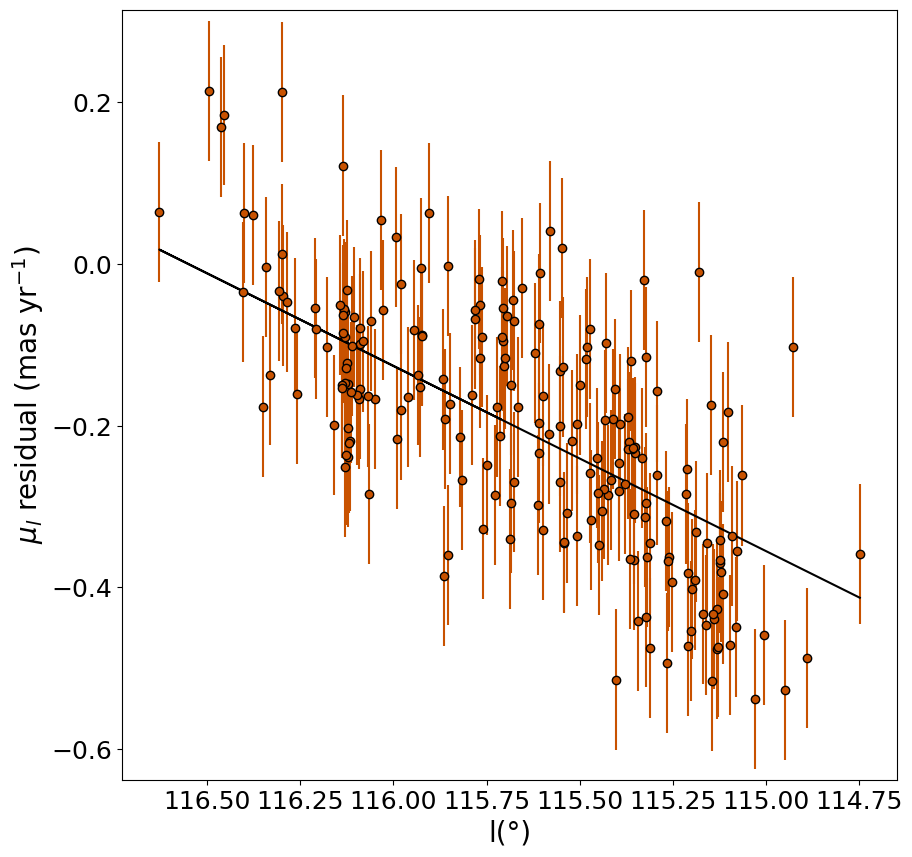}
    \caption{Linear fit between the residual proper motion in Galactic longitude (after subtraction from the Galactic rotation model from \citealt{Eilers2019}) and Galactic longitude for the individual members of Cluster 6, corresponding to an expansion age of $25 \pm 1$ Myr.}
    \label{ExpansionPlot}
\end{figure}

We have determined whether our clusters were expanding by fitting a linear relationship between position and proper motion in both the $l$ and $b$ directions. To that end, we have performed a MC simulation with 1000 iterations, randomly sampling the individual stellar proper motions and distances of each group within their uncertainties. We chose the expansion gradients as the median values, along with the 16th and 84th percentiles as the lower and upper error bars on these estimates, respectively. For the Galactic longitude direction, we removed the contribution in the proper motions due to Galactic rotation using the model from \citet{Eilers2019} (thus, we also randomly sampled the parameters from this model as part of the MC simulation). Subsequently, we inverted these expansion gradients into expansion ages for clusters with a significant expansion pattern, with the results displayed in Table \ref{ClusterExpansion}. An example of linear fit is given for Cluster 6 in Fig. \ref{ExpansionPlot}.

Table \ref{ClusterExpansion} shows that all our OCs (except Clusters 12 and 13) exhibit some expansion pattern in at least one direction, although they are generally more significant over Galactic longitude. Some OCs are expanding in both directions (Clusters 3-7, 10, 15, and 16), whereas others show evidence of contraction along one direction (Clusters 1, 2, 8 and 17). For the clusters whose expansion ages we were able to derive in both directions, there are cases of mutual agreement (Clusters 3 and 6) but also cases where ages in $l$ and $b$ disagree significantly (Clusters 4, 7, 10, 15, and 16). The expansion ages we estimated for Cluster 6 are consistent with the isochronal ages of King 12, NGC 7788 and FSR 451 from \citet{HuntReffert2024}, and the ages we found are also consistent with Stock 17 and Teutsch 23 (for Cluster 9) and with NGC 7790 and Berkeley 58 (for Cluster 11; see Table \ref{OCsProperties}). Interestingly, the expansion age for Cluster 8 is aligned with NGC 7790 and Berkeley 58, suggesting its kinematic signature is dominated by its older population.

\subsection{Clustered and diffuse population}
\label{clusterediffuse}

Out of the 56\,379 SED-fitted OBA stars spanning the $6\times8$~deg$^2$ area, 1156 were found to be clustered, including 490 members of the four OCs related to Cas~OB5. If we consider these cluster members as defining the association, then Cas~OB5 can be characterised in terms of 3D position ($l$, $b$ and $\varpi$) and 2D velocity ($\mu_l$ and $\mu_b$), with $l \in$ [$114.5\degr$, $117.5\degr$], $b \in$ [$-1.5\degr$, $1.5\degr$],  $\varpi \in$ [0.3, 0.4] mas, $\mu_l \in$ [-4.2, -3] mas yr$^{-1}$ and $\mu_b \in$ [-1.4, -0.2] mas yr$^{-1}$. Selecting all the objects within these limits results in a diffuse population of 2647 SED-fitted OBA stars within Cas~OB5, compared to 490 stars in the clustered population, thus, it is shown to be $\sim$5.4 times larger.

This result motivated us to investigate how many objects might have been recently ejected from these clusters. To this end, we defined each cluster's extent as ellipses centred on their median Galactic coordinates, with widths and heights corresponding to 3$\sigma$ from their median $l$ and $b$ values. Subsequently, we performed a 2D linear traceback for the 2647 field stars, going back up to 3~Myr (the effects of the Galactic potential make this approximation inaccurate beyond a few million years, e.g. \citealt{Fuchs}). We recorded every star that fell within the ellipse and whose parallax was consistent with being part of the cluster (i.e.\ within 1$\sigma$ from the median $\varpi$), resulting in 197, 182, 81, and 37 stars that could have been ejected from these four clusters, respectively. They add up to a total of 497 stars belonging to the `extended clustered population',  which is about the same number as in the clustered population.

We also estimated how likely it is for stars from the diffuse population to have been projected as part of the clustered population due to random motion. To that end, we kept the same Galactic coordinates for the diffuse population and uniformly randomised their proper motions (PMs) and parallaxes within the above defined borders of Cas~OB5. We performed a MC simulation with 1000 iterations, recording each time that a randomly assigned star fell inside a cluster. This resulted in 782, 258, and 1 stars, respectively, with a probability above 30, 40, and 50 \% to have been ejected from these clusters because of random motion. If we were to choose the threshold of 30 \%, this would imply that about one-third of this extended clustered population ($\sim$150 stars) could actually be attributed to random motion and the rest ($\sim$350 stars) is likely to represent real ejections.

In summary, if we combine the clustered population with the reliable `extended clustered population', this makes a population of $\sim$840 clustered stars in Cas~OB5. This can be contrasted with a diffuse population of $\sim$2297 stars. Therefore, the size of the diffuse population is reduced to a factor of $\sim$2.7 times bigger than the clustered population.

\section{Discussion}
\label{discussion}

\subsection{Limitations of clustering algorithms}
HDBSCAN has proven its capability to identify kinematically-coherent stellar groups of various densities, making it suitable for discovering OCs. Nevertheless, our work has shown that seemingly similar groupings (concentrated both in positional and velocity space) can in fact correspond to populations of unambiguously distinct ages. This is because the clustering method relies on the assumption that a group of stars, close in 3D position and 2D velocity space, should be related, whilst the actual stellar distribution can be more complex (especially with respect to age).

Consequently, in addition to complementing HDBSCAN with a method to remove false positives, it is crucial to investigate clusters individually to assess reliably their age and distinguish older from younger groups. Thus, this would require a careful examination of their colour-magnitude diagram and the exploitation of spectroscopic observations (as we show in Appendix~\ref{app:ngc}). 

In our analysis, Cluster~8 is barely separated from Cluster~11 (see Fig.~\ref{GalCoordClusters} and Fig.~\ref{GalVelocityClusters}); they were only identified as different groups because our clustering method gave more weight to the distance parameter compared with \citet{HuntReffert2024}. While we used the stellar distances for the clustering analysis, \citet{HuntReffert2024} used directly the \textit{Gaia} parallaxes. However, our clustering method identifies overlapping OCs and thus has the advantage of relating seemingly distinct populations of stars. The most noteworthy examples of this (as highlighted in Appendix~\ref{app:ngc}) are NGC~7790 and Berkeley~58, which form a single sequence, similarly to what was observed in the work of \citet{MaTur24}.

\subsection{Bound open clusters or unbound subgroups}

Many recent studies investigating OB associations with \textit{Gaia} data have focused on parsec scales, detecting compact subgroups within them  \citep[e.g.][]{WrightMamajek2018,Squicciarini2021,MiretRoig2022,Ratzenbock2023,Szilagyi2023}. The term `cluster' has sometimes been used to refer to them, albeit most of them are not gravitationally bound and just happen to extend over a similar scale as OCs. 

In principle, since we applied a clustering algorithm similar to those used in these studies, it is important to further analyse these groups to unveil their true nature. This is particularly relevant because OB associations are thought to be composed of subgroups of various ages \citep{Wright2020}. For instance, \citet{WrightMamajek2018} utilised the 3D velocity dispersion of the subgroups of the Sco-Cen association to determine their virial state and concluded they were gravitationally unbound. 

Nonetheless, in our case, our identified groups share many members with well-known compact OCs, whose bound nature was recently  confirmed by \citet{HuntReffert2024}. We can thereby designate them as OCs  belonging to the wider Cas~OB5 association. 

\subsection{Galactic structure}
\label{galstruct}

We have identified OCs strongly overlapping with King 12 and NGC~7790 and (to a lesser extent) NGC~7788, three OCs that are thought to be part of Cas~OB5 (e.g. \citealt{Kusuno2013}).  We have been able to confirm this information here thanks to their age and distance (see Table \ref{OCsProperties}). Furthermore, given their similar properties, we argue that Stock~17, FSR~451 and Berkeley~58 are probably also related to Cas~OB5. The compact clusters identified with our Cluster~9 (Teutsch~23, King~21) could represent an extension to the north.

Cas~OB5 is believed to intercept the Perseus arm \citep{DelaFuenteMarcos2009,MarcoNegueruela2016}. Recent investigations have shown that this arm is situated $\sim$2.5\,--\,3~kpc away at this Galactic longitude  \citep[e.g.][]{Reid2019}. In light of \textit{Gaia} DR3 data, and as highlighted in \citet{SaltovetsMcSwain2024}, this is consistent with the location of Cas~OB5.

\begin{figure*} 
\includegraphics[scale =0.27]{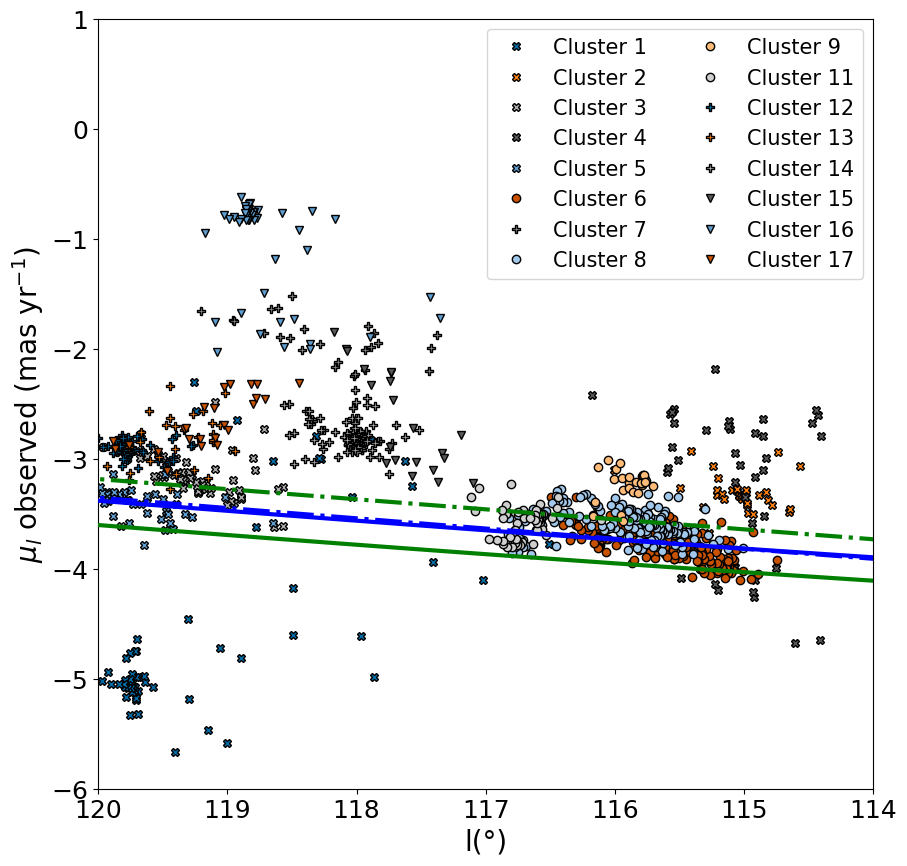} 
\includegraphics[scale = 0.27]{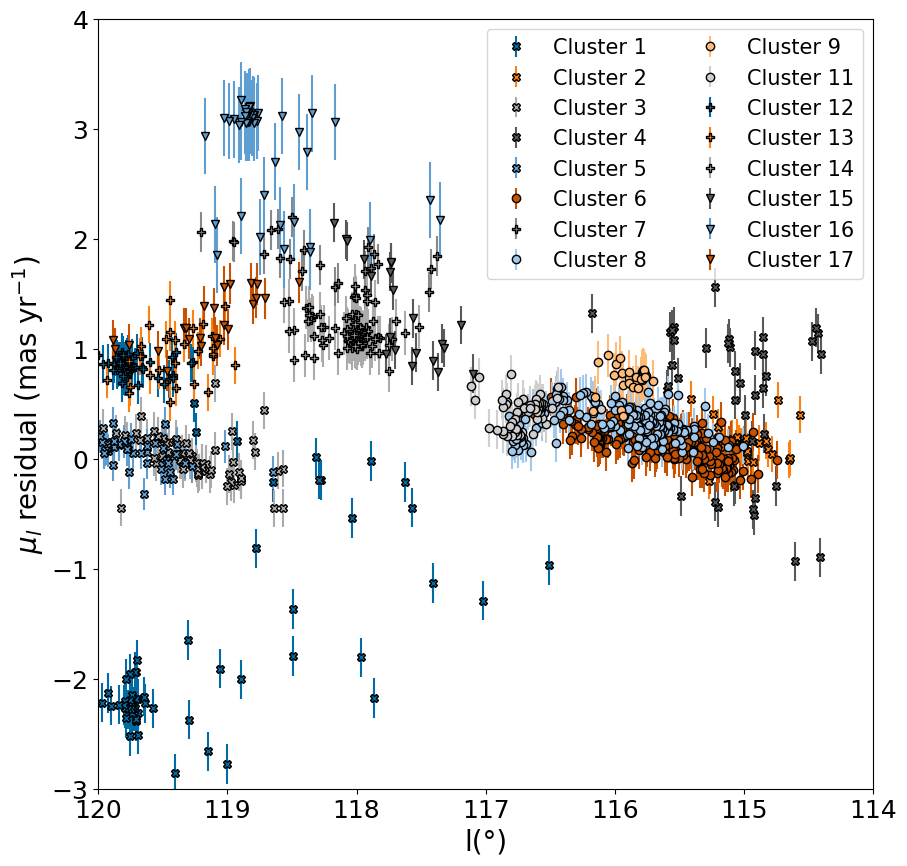}
\includegraphics[scale = 0.27]{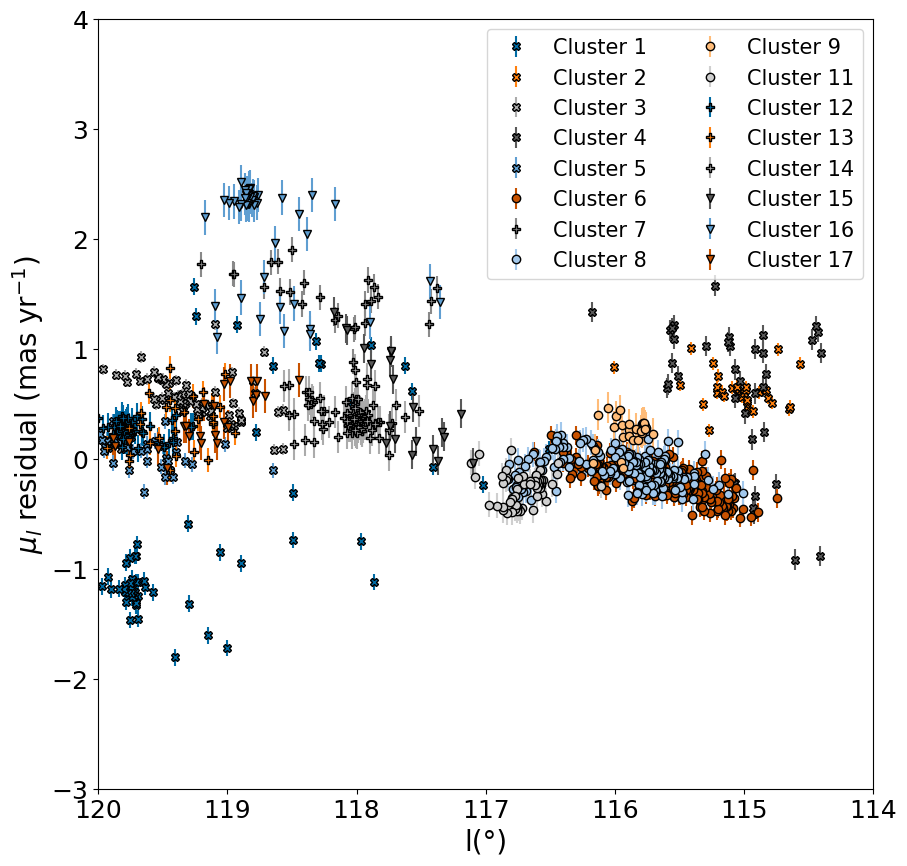}
\caption{Galactic longitude as a function of proper motions in Galactic longitude for our 17 identified clusters. The left panel shows observed PMs, with the blue and green lines respectively representing the rotation curves at a distance of 2.5 and 3 kpc (from \citealt{Bovy2017} as solid lines and from \citealt{Eilers2019} as dash-dotted lines). The middle panel displays the residual PMs after subtraction from the Galactic rotation models by \citet{Bovy2017}, and the right panel is identical but uses the model from \citet{Eilers2019}.}
\label{ProperMotRotStructure}

\end{figure*}

In spite of their different ages, our OCs related to Cas~OB5 are kinematically indistinguishable. We have shown their PMs as a function of their position in Galactic longitude in Fig.~\ref{ProperMotRotStructure}. We have used two different Galactic rotation models to visualise their residual PMs (by \citealt{Bovy2017}, where we used the velocities of the Sun with respect to the Local Standard of Rest from \citealt{Schronrich2010}, and by \citealt{Eilers2019}). In both cases, their residual PMs are close to zero, implying that the linear correlation that they shape is mostly a product of Galactic rotation, but also that the older OCs exhibit the same kinematic signature as the younger ones. 

Therefore, even though they overlap strongly in positional and velocity space with the younger clusters, NGC~7790 and Berkeley~58 cannot serve as tracers of the spiral arms, since a spectroscopic investigation has revealed that they are significantly older than the other OCs related to Cas~OB5. This would imply that Cas~OB5 is actually composed of a younger association in the foreground and an older association in the background ($\sim$300 pc more distant), with both having very similar proper motions, as we discuss in Appendix \ref{app:ngc}.

\subsection{Spatial scale of star formation}

As specified in Section \ref{clusterediffuse}, we  delimited Cas~OB5 through its four constituent and dynamically coherent OCs, from Cluster~6 at a median SED-fitted distance of 2.82~kpc to Cluster~11 at a median SED-fitted distance of 3.35~kpc. Given that Clusters 8 and 9 are located at median distances of 2.97 and 3.04~kpc, this implies a separation of about $\sim$150, 70 and 310 pc between the closest and the most distant OC. However, since Cluster~11 is older, it probably did not originate from the same star formation episode as Clusters~6, 8, and 9 (consistent with Cluster~11 also being more distant). If we consider only these three clusters, they can see that they extend over a distance of $\sim$200 pc along the line of sight.

As defined by these clusters, Cas~OB5 spans $\sim3\times2$~deg$^2$ across the plane of the sky. At a distance of $\sim$3~kpc, it is equivalent to $\sim150\times100$~pc$^2$. Therefore, this complex is consistent with a size of a few hundreds of parsecs in 3D positional space. Our findings, combined with previous estimations, suggest that Cas~OB5 has an age of about 10\,--\,20~Myr (which implies that Cas~OB5 cannot be the powering source of the H\,{\sc ii} region Sh2-173, contrary to what was claimed in the work of \citealt{Cichowolski2009}). Fully decoupled from their natal cloud, the young OCs from Cas~OB5 follow Galactic rotation and trace the position of the Perseus arm (see Fig.~\ref{ProperMotRotStructure}), with an extent that is consistent with the thickness of the spiral arm. Provided that Cas~OB5 expanded from its initial configuration, we observe star formation across three scales: from the smaller, compact OCs listed in \citet{HuntReffer2021} ($\sim$ pc) to the larger OCs identified by our method ($\sim$ tens of pc) and all the way up to the whole Cas~OB5 complex ($\sim$ hundreds of pc).

\section{Conclusions}
\label{conclusion}

In this paper, we  investigate the region around the Cas~OB5 association, characterising its higher mass stars (earlier than A5) through a SED fitter. We grouped these stars using HDBSCAN and identified seventeen reliable open clusters, four of which we found to be related to Cas~OB5, having strong overlaps with well-known compact clusters in the area. Most of these clusters are expanding in at least one direction, while a few of them are contracting, with expansion ages that are generally consistent with the isochronal ages of the known OCs they overlap with. We  also compared the clustered and the diffuse population of Cas~OB5 to estimate a spatial scale of a few tens to a few hundreds of pc for star formation in this region of the Perseus spiral arm. In forthcoming studies, we will investigate other associations in this arm.

Despite their proximity in positional and velocity space -- and despite their sharing a common kinematic signature (as evidenced by Galactic rotation models) -- the constituent OCs of Cas~OB5 exhibit diverse ages. We  confirmed this by using spectroscopic observations to obtain a robust age estimate for NGC~7788 (10\,--\,15 Myr) and NGC~7790 ($110\pm15\:$Myr). Consequently, we find that NGC~7790 and Berkeley~58 cannot serve as tracers of the Perseus spiral arm, even though they intercept it.

This work has  demonstrated that whilst modern clustering algorithms constitute a robust method to identify stellar groups and unravel Galactic structure, they should be combined with accurate spectroscopic data in order to reliably determine  their ages and establish tracers of the spiral arms.

Upcoming spectroscopic surveys such as WEAVE and 4MOST, combined with future \textit{Gaia} data releases, are therefore timely. Notably, the upcoming SCIP survey from WEAVE will provide spectroscopy for more than $\sim200\,000$ OBA stars across the Galactic plane \citep{WEAVE}. These luminous targets are potential members of stellar clusters and associations,  with the potential to improve   constraints on their properties. This will offer a better understanding of the structure and dynamics of the Milky Way, potentially in more distant and extinguished regions. 

\section*{Data availability}

The catalogue of OBA stars, the cluster properties, as well as the cluster members have been uploaded to Vizier at the time of publication.

\begin{acknowledgements} 

This research is partially supported by the Spanish
Government Ministerio de Ciencia, Innovaci\'on y Universidades and Agencia Estatal de Investigación (MCIU/AEI/10.130 39/501 100 011 033/FEDER, UE) under grants PID2021-122397NB-C21/C22 and Severo Ochoa Programme 2020-2024 (CEX2019-000920-S).
It is also supported by MCIU with funding from the European Union NextGenerationEU and Generalitat Valenciana in the call Programa de Planes Complementarios
de I+D+i (PRTR 2022), project HIAMAS, reference ASFAE/2022/017, and NextGeneration EU/PRTR and MIU (UNI/551/2021) through grant Margarita Salas-ULL.

 This paper makes uses of data processed by the Gaia Data Processing and Analysis Consortium (DPAC, https://www.cosmos.esa.int/web/gaia/dpac/consortium) and obtained by the Gaia mission from the European Space Agency (ESA) (https://www.cosmos.esa.int/gaia), as well as the INT Galactic Plane Survey (IGAPS) from the Isaac Newton Telescope (INT) operated in the Spanish Observatorio del Roque de los Muchachos. Data were also based on the Two Micron All Star Survey, which is a combined mission of the Infrared Processing and Analysis Center/California Institute of Technology and the University of Massachusetts.

This work also used \textit{TOPCAT} \citep{Topcat}, Astropy \citep{Astropy}, and the Vizier and SIMBAD databases (both operated at CDS, Strasbourg, France). It has also profitted from the WEBDA database, operated at the Department of Theoretical Physics and Astrophysics of the Masaryk University.

The appendix is partially based on data obtained at the  INT and WHT telescopes, operated on the island of La Palma by the Isaac Newton Group in the Spanish Observatorio del Roque de Los Muchachos of the Instituto de Astrof{\'\i}sica de Canarias.

\end{acknowledgements}


\bibliographystyle{aa}
\bibliography{Bibliography} 



\begin{appendix} 

\section{Parameters of NGC~7788, NGC~7790 and Berkeley~58}
\label{app:ngc}

Probably the most striking result of the analysis presented in this paper is the presence, in the area of Cas~OB5, of two distinct, spatially-overlapping populations, with very similar astrometric parameters, but clearly separate ages.  To evaluate the reality of this separation, we need to obtain better age estimates. For this purpose, we concentrate on two clusters that are very close in the sky, but belong to the two populations, NGC~7788 and NGC~7790. An overview of the area has recently been presented by \citet{MaTur24}. These authors reach two main conclusions: NGC~7790 and the nearby cluster Berkeley~58 have almost exactly the same parameters (proper motions, parallax and age), while NGC~7788 is slightly in the foreground and much younger. In fact, when properly dereddened the CMDs of NGC~7790 and Berkeley~58 can be combined into a single sequence, and the classical Cepheid CG~Cas, associated with Berkeley~58, has similar properties to those of the three Cepheids in NGC~7790 \citep{MaTur24}. This is fully consistent with the results presented here, as our algorithm cannot distinguish between NGC~7790 and Berkeley~58, and assigns members of both to our Cluster~8.

NGC~7790 is a well-studied cluster, as it hosts three classical Cepheids, CE~Cas~A, CE~Cas~B, and CF~Cas, the highest number of Cepheids known in a Milky Way cluster. The most recent work by \citet{Majaess13}, based on photometric analysis, concluded that its distance is $3.40\pm0.15$~kpc, in perfect agreement with our distance for Cluster 11, but incompatible with the 3.1~kpc given by \citet{HuntReffert2024}. They also found an average reddening $E(B-V)=0.52$, with standard deviation $\sigma=0.05$~mag. Using isochrone fits, \citet{Majaess13} determine an age $\log\,\tau=8.0\pm0.1$ (100~Myr). \citet{HuntReffert2023}, using an automated algorithm to fit \textit{Gaia} photometry, find an age $67^{+36}_{-25}$~Myr and an extinction $A_V=1.70^{+0.16}_{-0.22}$~mag, which is compatible with the reddening found by \citet[and references therein]{MaTur24} if a standard reddening law is assumed.

NGC~7788 has received much less attention. The only dedicated study is that of \citet{Becker65}, who estimated a turn-off photometric type of b4. More recently, \citet{Davidge2012} used Sloan photometry to derive a distance modulus $\mu=12.07$ ($d=2.6\:$kpc) and $E(B-V)=0.55$. The age could not be constrained because the turn-off was not observed. \citet{Glushkova13} used Johnson/Kron photometry (without $U$) to calculate an age of 160~Myr and a reddening around 0.5~mag. The automated algorithm of \citet{HuntReffert2023} gives an age of $20^{+9}_{-8}$~Myr for a \textit{Gaia} distance $d=2.7\:$kpc, and an extinction $A_V=1.55^{+0.18}_{-0.19}$~mag, which is broadly compatible with the reddening values.

Here, we will complement the photometry with extensive spectroscopic data for NGC~7790 and NGC~7788. In what follows, we will use the numbering system of the WEBDA database\footnote{At \url{https://webda.physics.muni.cz/}} for stars in the clusters.

\subsection{Observations}

Observations of stars in the field of NGC~7790 were obtained on two different runs. The brightest objects  were observed on the night of August 22, 2007, with the blue arm of the ISIS spectrograph, mounted on the 4.2-m William Herschel Telescope (WHT), in La Palma (Spain). The instrument was equipped with the EEV12 CCD. We used grating R1200B (nominal dispersion of 0.23\,\AA/pixel) with a $1\farcs2$ slit, which gives a resolving power $R\approx4\,000$. Some fainter objects that fell by chance inside the slit also produced useful spectra.

A larger set of stars was observed with the same instrument on November 12, 2008. This time the two arms of the instrument were used. The blue arm was equipped with the EEV12 CCD and the R300B grating. The red arm used the Red+ CCD and the R316R grating. With a $1\farcs0$ slit, both setups give resolving powers somewhat above  $R\approx1\,000$. 

Stars in the field of NGC~7788 were observed on the night of September 11, 2006 with the Intermediate Dispersion Spectrograph (IDS) mounted on the 2.5~m Isaac Newton Telescope (INT), also in La Palma (Spain). The instrument was equipped with the 235-mm camera and EEV10 CCD. The R632V grating together with a $1\farcs0$ slit gives a resolving power $R\approx2\,500$. Additionally, one star (71  = LS~I~$+61\degr$106) had been observed during a different run with the INT+IDS, in July 2002. On that occasion, the instrument was configured with the
R1200Y grating and Tek\#4 CCD, a combination that gives a resolving power $R\approx3\,000$.
All the spectra were reduced  with  the \textit{Starlink} packages {\sc CCDPACK} \citep{draper2000} and {\sc FIGARO} \citep{shortridge93} and analysed using {\sc FIGARO} and {\sc DIPSO} \citep{howarth04}.

\subsection{Stellar content}

Our observations targeted all the brightest blue stars in both clusters, except for BD~$+60\degr$2644, the brightest star in NGC~7788. This object was recently classified as B1\,III, based on a high-resolution spectrum by \citet{DeBurgos2023}. 

Spectral classification has been carried out by application of classical criteria. The ISIS/R1200B and IDS spectra were directly compared to the new set of standards from \citet{Negueruela24}\footnote{Available at \url{https://astroplus.ua.es/mkbtypestds/}}, which have similar resolution. Several stars in common with the lower resolution ISIS/R300B dataset were used to ensure consistency.

   \begin{figure}
   \centering
    \resizebox{\columnwidth}{!}{\includegraphics[clip]{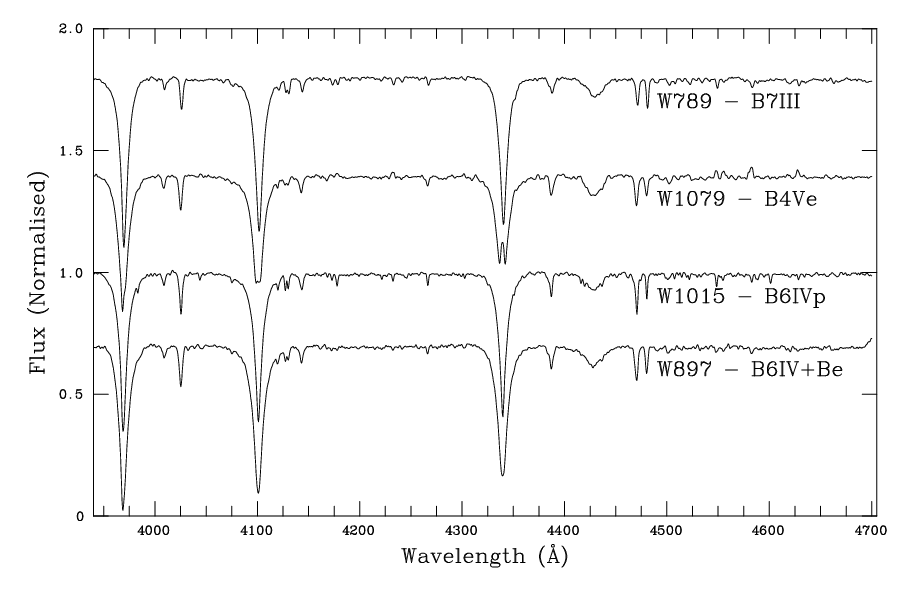}}
   \caption{Intermediate-resolution spectra of the brightest blue stars in NGC~7790. Note that W1079 and W1015 are not astrometric members. }
              \label{n7790bright}
    \end{figure}

Illustrative spectra obtained with ISIS/R1200B are displayed in Fig.~\ref{n7790bright}. Three stars that fall on the cluster's photometric sequence for which we have spectra are not picked as members by \citet{HuntReffert2024}. Star~1025 has totally unreliable astrometric parameters, as it has a RUWE=25.7. Its spectrum indicates B9\,V, which is not incompatible with its position in the CMD, if it is a binary. Star 1079 is the third-brightest blue star in the cluster sequence. The star is in the middle of the cluster, its proper motions are compatible with membership, and its spectral type B4\,Ve is fully compatible with its position in the CMD. However, its parallax $\varpi=0.38\pm0.02$~mas is incompatible with membership. If it is a cluster member, then it would be a blue straggler. Star~1015 is the second-brightest blue object in the cluster sequence (only 0.01~mag brighter than 1079). It is also in the centre of the cluster, and its spectral type B6\,IVp is fully compatible with its position in the CMD. Its parallax is typical of members, but its pmRA is divergent. 

In addition, we observed two stars in the halo of NGC~7790 that had previously been considered possible blue stragglers. None of them are members based on their \textit{Gaia} data. Star 1360 is QX~Cas, catalogued as a 6.0-d eclipsing binary, although its eclipses seem to have stopped \citep{guinan12}. The system has been reported to include two B-type stars, but our spectrum only shows a B0.5\,V spectrum with very narrow lines. Star 586 (LS~I~$+60\degr$63) is catalogued as a Be star, but this is not evident in our data, which include a red spectrum centred on H$\alpha$. We classify it as a B1.5\,V star.

   \begin{figure}
   \centering
    \resizebox{\columnwidth}{!}{\includegraphics[clip]{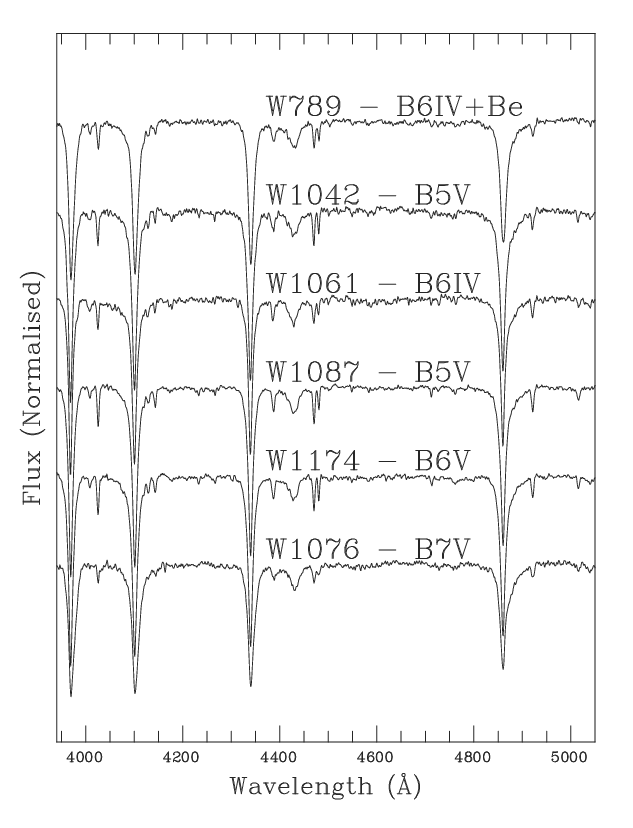}}
   \caption{Low-resolution spectra for stars around the turn-off of NGC~7790. }
              \label{n7790faint}%
    \end{figure}

Low-resolution ISIS/R300B spectra of fainter cluster members are shown in Fig.~\ref{n7790faint}, together with the spectrum of W789, to be compared to that in Fig.~\ref{n7790bright}. The spectral classifications for astrometric members are listed in Table~\ref{tab:members7790}, while data for objects not listed as members by \citet{HuntReffert2024} are given in Table~\ref{nonmembers}.

In the case of NGC~7788, all the brightest members according to \citet{HuntReffert2024} are in the cluster halo and were observed. However, some of the stars in the cluster core are not selected as members. As in the case of the two bright stars in the core of NGC~7790, the deviations in astrometric parameters are not large, and their position in the CMD is perfectly suited to their spectral type, suggesting that the deviations from the cluster mean values are caused by internal dynamical processes. In Table~\ref{tab:members7788}, we list the spectral types of all the stars observed, plus BD~$+60\degr$2644. with the astrometric non-members indicated in the last column.

Star 10 (TYC 4281-2414-1), which shows very broad lines, is a photometric variable and eclipsing binary candidate in \textit{Gaia} \citep{Mowlavi23}. Star 78 (LS~I~$+61\degr$108) also has broad lines, and is a strong binary candidate. The most interesting peculiar star is 95 (LS~I~$+61\degr$103), which displays very strong \ion{He}{i} lines and is a very strong candidate to be a magnetic star \citep[cf. e.g.][]{Jarvinen18}.

\begin{table*}
\caption{Spectral classification for astrometric members of NGC~7790, ordered by brightness. }           
\label{tab:members7790}     
\centering                          
\begin{tabular}{l l c c l}        
\hline\hline                 
\noalign{\smallskip}
ID& Name & Spectral & G & Notes\\   
 & &Type &(mag) &\\
 \noalign{\smallskip}
\hline                       
\noalign{\smallskip}
789 &           GSC 04281$-$01780 &B7\,III & 12.00 &Brightest blue star\\
897 &           TYC 4281-1626-1  &B6\,IV  & 12.66      & SB2, companion is Be \\
1042 &          GSC 04281--02198 & B5\,V   & 12.93      & Wrongly given as RS CVn in SIMBAD\\        
1061 &          GSC 04281--02486&   B6\,V  & 12.99      & \\              
1087  &         GSC 04281--02340 & B5\,V      & 13.01 \\                  
1174  &         GSC 04281--01970  & B6\,V      & 13.21\\  
1076 &  & B7\,V        & 13.36  \\      
1025 &          GSC 04281--02132 & B9\,V &  13.74& Membership uncertain\\
913 &           TIC 326815440  &B8\,V        & 13.97 &  \\
1187 &  TIC 326815628 & B8.5\,V        & 14.40  &  \\ 
\noalign{\smallskip}
\hline                                   
\end{tabular}
\end{table*}

\begin{table*}
\caption{Spectral classification and \textit{Gaia} DR3 data for astrometric non-members in the field of NGC~7790, for which we have spectra. }         
\label{nonmembers}    
\centering                         
\begin{tabular}{l l c c  c c c}    
\hline\hline                 
\noalign{\smallskip}
ID& Name & Spectral & G & pmRA & pmDec & $\varpi$\\   
 & & Type & (mag)& (mas/a)& (mas/a) & (mas)\\
 \noalign{\smallskip}
\hline                       
\noalign{\smallskip}
1079 &          GSC 04281--02486  & B4\,Ve & 12.58 & $-3.33\pm0.02$ &$-1.82\pm0.02$& $0.38\pm0.02$\\
1015 &          TYC 4281-1704-1 & B6\,IVp & 12.57&  $-2.92\pm0.01$ &$-1.91\pm0.01$&$0.29\pm0.01$      \\ 
\noalign{\smallskip}
\hline
\noalign{\smallskip}
1360  & QX~Cas & B0.5\,V  &10.12   &$-2.93\pm0.02$ & $-1.83\pm0.02$ & $0.27\pm0.02$     \\      
586 & LS~I~$+60\degr$63 &B1.5\,V & 10.97& $-3.05\pm0.02$ & $-1.83\pm0.02$ & $0.31\pm0.02$\\
\noalign{\smallskip}
\hline
\noalign{\smallskip}
615 &   GSC 04281--02439 &A2\,V& 13.11 & $-1.73\pm0.01$ & $-1.65\pm0.01$ &$0.38\pm0.01$\\
\noalign{\smallskip}
\hline                                   
\end{tabular}
\end{table*}

   \begin{figure}
   \centering
    \resizebox{\columnwidth}{!}{\includegraphics[clip]{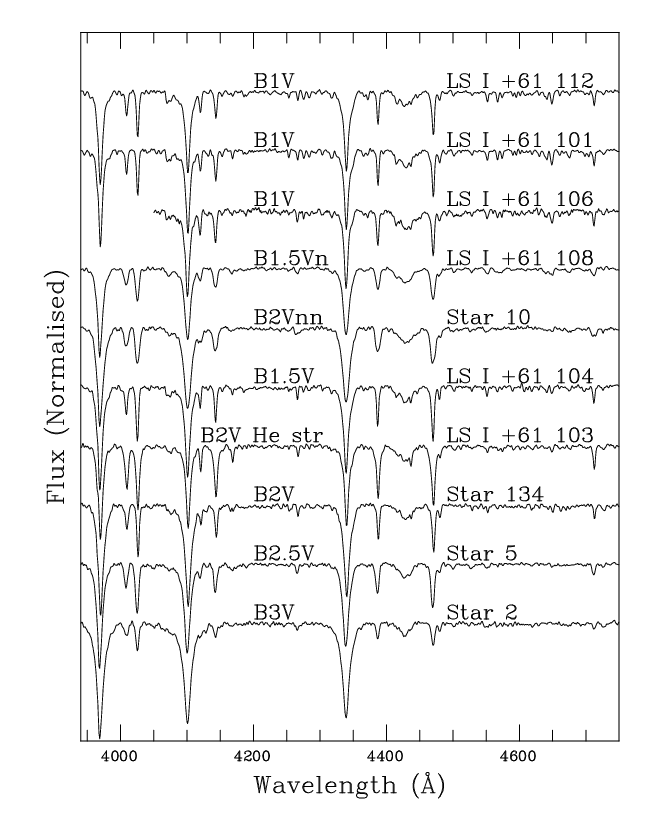}}
   \caption{Classification spectra for the brightest stars in NGC~7788, except for BD~$+60\degr$2644. Note the much stronger \ion{He}{i} lines in the spectrum of  LS~I~$+61\degr$103. }
              \label{n7788stars}%
    \end{figure}

\begin{table}
\caption{Spectral classification for stars in NGC~7788. }            
\label{tab:members7788}      
\centering                          
\begin{tabular}{l l c l}       
\hline\hline                
\noalign{\smallskip}
ID& Other name & Spectral & Notes\\   
 && type &\\
 \noalign{\smallskip}
\hline                       
\noalign{\smallskip}
1 & BD~$+60\degr$2644 & B1\,III & ANM \\
2  &  UCAC4 758--080753 & B3\,V        & ANM \\           
5  & UCAC4 758--080775 & B2.5\,V      &         \\        
10  &   TYC 4281-2414-1   & B2\,Vnn        & Likely binary, ANM\\       
68  &   LS~I~$+61\degr$104        & B1.5\,V        &              \\
71  & LS~I~$+61\degr$106  & B1\,V      & \\               
78  & LS~I~$+61\degr$108 & B1.5\,Vn       & Possibly binary\\  
95  & LS~I~$+61\degr$103   & B2\,V       &      He-strong\\
134  & TYC 4281-386-1 & B2\,V        &  \\      
136 &   LS~I~$+61\degr$112        & B1\,V              & Brightest dwarf \\
153  &  LS~I~$+61\degr$101                & B1\,V        &   \\
\noalign{\smallskip}
\hline                                   
\end{tabular}
\begin{minipage}{\linewidth} 
\vspace{0.25cm}
Note: ANM stands for astrometric non-member
\end{minipage}
\end{table}

\subsection{Cluster parameters}
\subsubsection{NGC~7790}

\begin{figure}
 \centering
  \resizebox{\columnwidth}{!}{\includegraphics[clip]{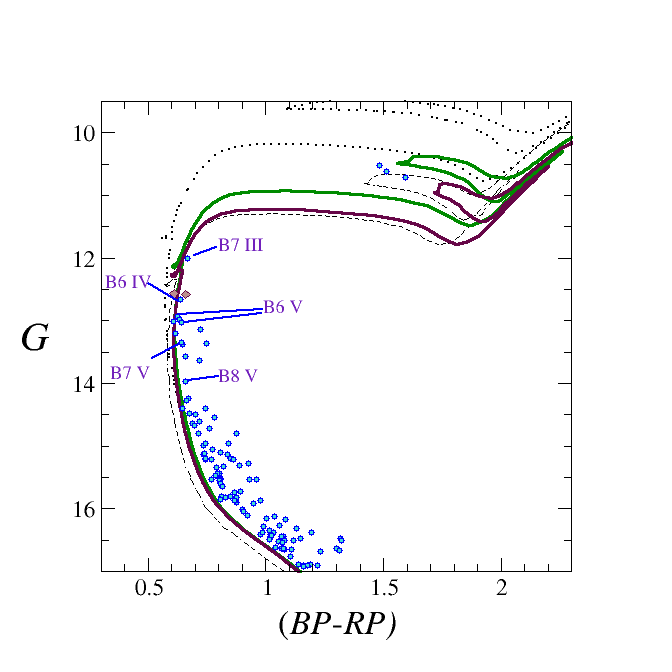}}
 \caption{\textit{Gaia} DR3 CMD for stars selected as members of NGC~7790 by \citet{HuntReffert2024}, plus two bright photometric members with divergent astrometric parameters (green diamonds). Several {\sc parsec} isochrones \citep{bressan12} displaced by the \textit{Gaia} $DM=12.6$ are shown. The solid lines are solar metallicity isochrones; the green line is a 105~Myr isochrone affected by $A_V=1.70$~mag, while the maroon one is a 130~Myr solar isochrone affected by $A_V=1.65$~mag, As reference, a 140~Myr $Z=0.010$ isochrone with $A_V=1.6$~mag is shown as a dashed line. The best solution found by \citet{HuntReffert2024} is shown as a dotted line. }
  \label{fig:n7790_fit}%
 \end{figure}

The \textit{Gaia} CMD for NGC~7790 is shown in Fig.~\ref{fig:n7790_fit}, with representative spectral types indicated. Star 789 (B7\,III) is the only blue star clearly evolved away from the main sequence. All the other bright stars are completely consistent with a turn-off at B5/6. The spectral types displayed indicate an age marginally younger than that of the Pleiades \citep[see][appendix C]{Negueruela24}, where there are also B6\,IV stars, but no dwarfs earlier than B7. Well-populated clusters with similar distributions of spectral types are NGC~6664 \citep{Alonso20} or NGC~6067 \citep{Alonso17}, which have ages just below 100~Myr.

The cluster parameters preferred by \citet{HuntReffert2024} very clearly do not provide a good fit for the cluster CMD (dotted line in Fig.~\ref{fig:n7790_fit}). The 105~Myr {\sc parsec} isochrone gives a good fit to the blue stars, but its blue loop does not quite reach the position of the Cepheids. Older isochrones have even shorter loops (we plot the 130~Myr one as an example), while younger isochrones, which have more extended blue loops, would have more luminous Cepheids. 

Isochrones of lower metallicity (one is shown as an example) have blue loops reaching the position of the Cepheids. The metallicity of NGC~7790, however, is very firmly determined as solar, based precisely on spectroscopic analysis of the Cepheids \citep[and references therein]{Groene18}. It is possible that the {\sc parsec} isochrones somewhat underestimate the extent of the blue loops at solar metallicity, although the issue may be also related to the reddening procedure applied to the isochrones by the {\sc cmd} interface. 

   \begin{figure}
   \centering
    \resizebox{\columnwidth}{!}{\includegraphics[clip]{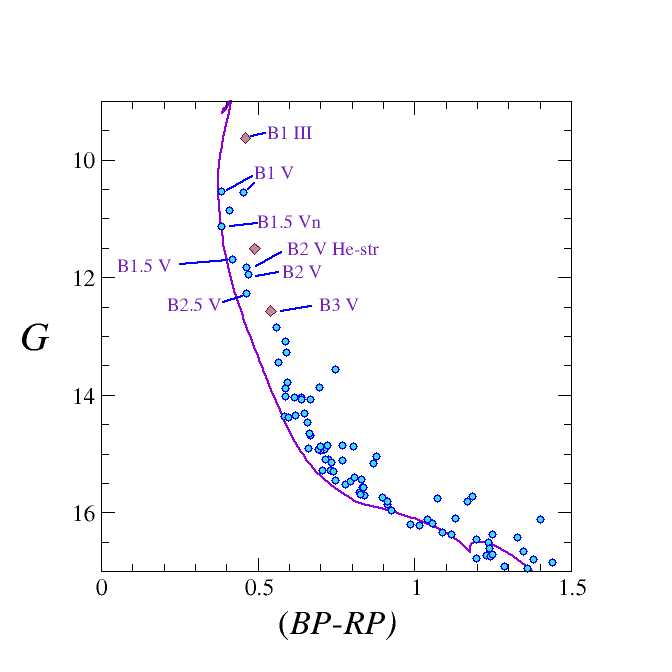}}
   \caption{\textit{Gaia} DR3 CMD for stars selected as members of NGC~7788 by \citet{HuntReffert2024}, plus three bright photometric members with divergent astrometric parameters (green diamonds) for which we have spectra. An illustrative 14~Myr solar-metallicity {\sc parsec} isochrone is shown, displaced by the \textit{Gaia} $DM=12.3$, and affected by $A_{V}=1.55$, after \citet{HuntReffert2023}.}
              \label{cmd7788}%
    \end{figure}
In any event, it seems clear that the age of NGC~7790 is not very different from 100~Myr. Trusting the isochrones, we would give a value of $110\pm15\:$Myr, in good agreement with the result of \citet{Majaess13}, and not far from the upper limit permitted by the errors in \citet{HuntReffert2024}. Since \citet{MaTur24} conclude that NGC~7790 and Berkeley~58 are twin clusters, the latter is expected to have the same parameters. The spectral types of the two early-B stars observed in the field seem far too early to allow the possibility that these two stars are blue stragglers ejected from the cluster (in the case of 586, the very direction of the proper motion prevents this possibility). They must be part of the diffuse Cas~OB5 population that extends between clusters.

NGC~7790 is the only open cluster known to host three classical Cepheids in the Milky Way. Nevertheless, the present analysis conclusively shows that this is not a massive cluster. Even if we assume that the astrometric non-members that fit the cluster CMD are dynamically disturbed members, there are less than 25 B-type members, and at best 15 stars in the B5\,--\,B7 range. This can be compared to the approximately 60 B-type members in NGC~6649 or NGC~6664, which contain only one Cepheid each \citep{Alonso20}. The current mass estimate of $1\,080\pm180\:\mathrm{M}_{\sun}$ given by \citet{HuntReffert2024} is, in all likelihood, an overestimate, since their best age is significantly younger than our result. The true mass of the cluster is likely to be somewhat below their lower limit of $900\:\mathrm{M}_{\sun}$, while the estimated masses of NGC~6649\footnote{\citet{HuntReffert2024} estimate $\approx10\,000\:\mathrm{M}_{\sun}$ for NGC~6649, but this is based on their best-fit age of 27~Myr, which is much younger than the 63~Myr given by \citet{Alonso20}, and incompatible with the observed tun-off at B5\,V. Nevertheless, the mass of NGC~6649 is likely to be significantly higher than the $2\,600\:\mathrm{M}_{\sun}$ estimated by \citet{Alonso20}.} and NGC~6664 are around $3\,000\:\mathrm{M}_{\sun}$ \citep{Alonso20}.

\subsubsection{NGC~7788}

The \textit{Gaia} CMD for NGC~7788 is shown in Fig.~\ref{cmd7788}, with representative spectral types indicated. Star 1 (BD~$+60\degr$2644) is the only object that shows any sign of evolution. All the other members are dwarfs, and moderately close to the ZAMS. The presence of B1\,V stars places an upper limit on the age of the cluster close to 15~Ma. Stars around the turn-off are comparable to those in the much more massive $h$~Per cluster. If BD~$+60\degr$2644 is indeed a member, it would suggest an age older than 10~Myr. In any event, the cluster is decidedly younger than all previous estimates, although compatible with the lower error margin given by  \citet{HuntReffert2023}. The lack of evolved stars does not permit a better age determination. The mass of the cluster is unlikely to be very different from the $840\pm110\:\mathrm{M}_{\sun}$ given by \citet{HuntReffert2024}.

\subsection{Implications}

Despite their similar distances and proper motions, the ages of NGC~7788 and NGC~7790 are very different. The star formation processes that created these two clusters took place at least 80~Myr apart, a timespan that seems far too long for a single event. Though separated in the sky by only $17\arcmin$, which is equivalent to 12~pc at 3~kpc, their actual physical separation is much greater, as their distances differ by at least 300~pc, confirming that there is no physical connection between the two. The similarity in their proper motions can only be interpreted as indicating that their space velocities are predominantly determined by Galactic rotation. 

Despite this, the separation of the two clusters in astrometric space is not trivial. According to \citet{HuntReffert2024}, stars in NGC~7788 have an average $\varpi=0.34$~mas, with a standard deviation of $0.02$~mas, while those in NGC~7790 display average $\varpi=0.29$~mas, with a standard deviation of $0.03$~mas. Likewise, their average proper motions in both right ascension and declination are distinct, but their standard deviations are sufficiently large for the extreme values to overlap. When considering an isolated star, such as the early-B objects in the field of NGC~7790 (Table~\ref{nonmembers}), there is no compelling reason to assign them to either population, although their proper motions align better with those of NGC~7788, and they would probably be considered cluster members if found in its field. This would be consistent with the ages implied by their spectral types.

The most plausible conclusion of this analysis, when combined with the results presented in the main body of this paper, is that Cas~OB5 is a young association comprising a number of moderately small clusters and a large diffuse population, which is projected just in front (by about 300~pc) of an older association that shares very similar astrometric parameters. In a future work, we will present detailed studies of the clusters that, along with NGC~7788, make up this association.   

\section{Cluster parameters}

We  give the main parameters (position and velocity) of the 17 OCs identified in Section \ref{identificationclusters}, including a comparison with the overlapping OCs from \citet{HuntReffert2024}.

\begin{sidewaystable*}
    \centering
    \caption{Main spatial, astrometric, and comparison parameters for the 17 OCs identified in Section \ref{identificationclusters}. }\label{TabClus}
    \renewcommand{\arraystretch}{2.0}  

    \resizebox{\textwidth}{!}{
    \begin{tabular}{cccccccccccccccc} 
        \hline \hline
        OC & $N$ & $l$ (deg) & $b$ (deg) & $\varpi$ (mas) & $\mu_l$ (mas yr$^{-1}$) & $\mu_b$ (mas yr$^{-1}$) & $d$ (kpc) & $V_l$ (km s$^{-1}$) & $V_b$ (km s$^{-1}$) & $N_{\rm HR24}$ & OC (HR24) \\
        \hline
        1 & 58 & 119.67 & 1.70 & $0.59 \pm 0.02$ & $-4.98 \pm 0.01$ & $-0.79 \pm 0.01$ & $1.71^{+0.08}_{-0.07}$ & $-40.39 \pm 1.64$ &  $-6.65 \pm 0.20$ & 27 & King 1 \\
        2 & 33 & 115.02 & -2.92 & $0.50 \pm 0.02$ & $-3.33 \pm 0.01$ & $-0.28 \pm 0.01$ & $2.00^{+0.12}_{-0.11}$ & $-31.15 \pm 1.58$ &  $\-2.75 \pm 0.06$ & 20 & UBC 180 \\ 
        3 & 64 & 119.29 & 0.63 & $0.50 \pm 0.02$ & $-3.17 \pm 0.01$ & $-0.79 \pm 0.01$ & $2.02^{+0.13}_{-0.10}$ & $-30.42 \pm 1.49$ &  $-7.63 \pm 0.25$ & 21, 4 & Theia 2230, CWNU 1780 \\
        4 & 41 & 115.07 & -1.19 & $0.42 \pm 0.02$ & $-3.01 \pm 0.01$ & $-0.86 \pm 0.02$ & $2.39^{+0.17}_{-0.15}$ & $-34.61 \pm 2.2$ &  $-9.68 \pm 0.41$ & 1, 1 & UBC 1189, UBC 1602 \\ 
        5 & 42 & 119.74 & -0.59 & $0.39 \pm 0.02$ & $-3.37 \pm 0.01$ & $-0.90 \pm 0.01$ & $2.58^{+0.12}_{-0.11}$ & $-41.41 \pm 2.16$ &  $-10.99 \pm 0.43$ & 15, 4, 4 & Stock 20, NGC 103, Mayer 1 \\
        6 & 211 & 115.61 & 0.09 & $0.36 \pm 0.03$ & $-3.76 \pm 0.01$ & $-0.64 \pm 0.02$ & $2.81^{+0.20}_{-0.17}$ & $-50.18 \pm 3.31$ &  $-8.46 \pm 0.31$ & 30, 22, 4, 3 & King 12, Stock 17, NGC~7788, FSR 451 \\
        7 & 26 & 118.15 & -0.66 & $0.36 \pm 0.03$ & $-1.92 \pm 0.01$ & $-0.71 \pm 0.02$ & $2.88^{+0.27}_{-0.24}$ & $-26.37 \pm 2.18$ &  $-9.81 \pm 0.46$ & 1 & UBC 1196 \\
        8 & 188 & 115.81 & -1.11 & $0.35 \pm 0.03$ & $-3.61 \pm 0.01$ & $-1.15 \pm 0.02$ & $2.95^{+0.30}_{-0.22}$ & $-50.39 \pm 4.00$ &  $-16.15 \pm 1.05$ & 63, 13, 8, 7 & FSR 451, Berkeley~58, NGC~7790, NGC~7788 \\
        9 & 30 & 115.85 & 1.04 & $0.35 \pm 0.03$ & $-3.21 \pm 0.01$ & $-0.69 \pm 0.03$ & $3.01^{+0.36}_{-0.31}$ & $-45.39 \pm 4.88$ &  $-9.76 \pm 0.71$ & 6, 3, 2 & Teutsch 23, King 21, Negueruela 1 \\
        10 & 88 & 119.45 & 0.66 & $0.33 \pm 0.03$ & $-3.16 \pm 0.01$ & $-0.28 \pm 0.02$ & $3.16^{+0.30}_{-0.24}$ & $-46.82 \pm 3.76$ &  $-4.21 \pm 0.11$ & 8, 6 & UBC 407, SAI 4 \\
        11 & 61 & 116.66 & -1.03 & $0.31 \pm 0.02$ & $-3.55 \pm 0.01$ & $-1.04 \pm 0.02$ & $3.34^{+0.34}_{-0.27}$ & $-56.73 \pm 5.01$ & $-16.67 \pm 1.15$ & 27, 14, 1 & NGC~7790, Berkeley~58, UBC 406 \\
        12 & 73 & 119.78 & -1.39 & $0.31 \pm 0.03$ & $-2.91 \pm 0.01$ & $-0.78 \pm 0.02$ & $3.37^{+0.34}_{-0.30}$ & $-46.30 \pm 4.35$ &  $-12.46 \pm 0.82$ & 57 & NGC 103 \\
        13 & 29 & 119.41 & 0.39 & $0.30 \pm 0.03$ & $-2.83 \pm 0.01$ & $-0.47 \pm 0.02$ & $3.42^{+0.31}_{-0.30}$ & $-45.93 \pm 3.99$ &  $-7.04 \pm 0.28$ & 6, 1, 1 & HSC 947, UBC 407, SAI 4 \\
        14 & 119 & 118.01 & -1.30 & $0.29 \pm 0.03$ & $-2.82 \pm 0.01$ & $-0.45 \pm 0.02$ & $3.55^{+0.43}_{-0.34}$ & $-47.46 \pm 4.96$ &  $-7.55 \pm 0.40$ & 76 & King 13 \\
        15 & 21 & 117.72 & 2.06 & $0.29 \pm 0.03$ & $-2.78 \pm 0.01$ & $-0.35 \pm 0.02$ & $3.55 \pm 0.34$ & $-46.30 \pm 4.10$ &  $-5.92 \pm 0.23$ & 0 &  \\
        16 & 48 & 118.83 & -1.68 & $0.29 \pm 0.04$ & $-0.82 \pm 0.01$ & $-0.65 \pm 0.03$ & $3.59^{+0.44}_{-0.45}$ & $-13.66 \pm 1.55$ &  $-11.02 \pm 0.82$ & 26 & Berkeley 60 \\
        17 & 24 & 119.01 & 0.39 & $0.27 \pm 0.04$ & $-2.72 \pm 0.01$ & $-0.34 \pm 0.03$ & $3.83^{+0.49}_{-0.46}$ & $-49.93 \pm 5.65$ &  $-6.09 \pm 0.25$ & 5, 1 & Theia 3832, HSC 947 \\
        \hline
    \end{tabular}%
    }

    \vspace{0.5cm}
    \small Note: $N$ is the total number of stars in our clusters. We have also compared our cluster members with those from \citet{HuntReffert2024}, with $N_{\rm HR24}$ standing for how many cluster members crossmatch, ordered from the largest to the smallest correspondence.
\end{sidewaystable*}

\section{3D distribution of the clusters}
\label{3Ddis}
\begin{figure}
    \centering
    \includegraphics[scale = 0.6]{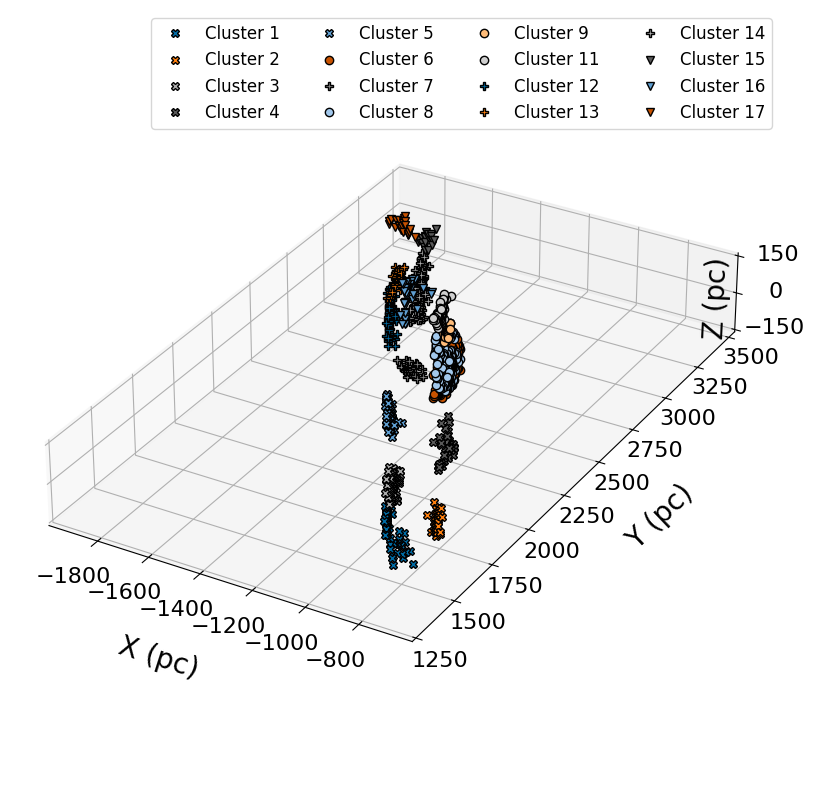}
    \caption{3D distribution in Cartesian coordinates of the 17 open clusters identified in Section \ref{identificationclusters}.}
    \label{3Dplotsclusters}
\end{figure}

We  display a 3D plot in Fig. \ref{3Dplotsclusters} of the 17 OCs in order to visualise their XYZ distribution. We observe an interesting trend of increasing Z values for decreasing X and increasing Y values. Unsurprisingly, most of the clusters are located at $Y \in$ [2.5 kpc, 3 kpc].

\end{appendix}
\end{document}